\documentclass[twocolumn,prb,aps,superscriptaddress,footinbib]{revtex4-2}
\PassOptionsToPackage{dvipsnames,svgnames,table}{xcolor}
\usepackage{array}
\usepackage{xr}
\usepackage{graphicx}
\usepackage{dcolumn}
\usepackage{color}
\usepackage{bm}
\usepackage{comment}
\usepackage{tabularray}
\usepackage{multirow}
\usepackage{amsmath,amssymb,amsfonts,bbold}
\usepackage{mathtools}
\usepackage[normalem]{ulem}
\usepackage{nameref}
\usepackage{setspace}
\usepackage{float}
\usepackage{ragged2e}
\usepackage{physics}

\usepackage[dvipsnames]{xcolor}

\usepackage[hidelinks]{hyperref}
\hypersetup{
  colorlinks   = true, %Colours links instead of ugly boxes
  urlcolor     = blue, %Colour for external hyperlinks
  linkcolor    = blue, %Colour of internal links
  citecolor   = blue %Colour of citations
}

\usepackage{booktabs}

\newcommand{\cre}[2]{{#1}^\dagger_{#2}}
\newcommand{\des}[2]{{#1}_{#2}}

\newcommand{\one}{\mathbb{1}}

\graphicspath{{fig/}}
 
\begin{document}
\title{Fragility of local moments against hybridization with flat bands}

\author{Max~Fischer}
\affiliation{Institut f{\"u}r Theoretische Physik und Astrophysik and W{\"u}rzburg-Dresden Cluster of Excellence ctd.qmat, Universit{\"a}t W{\"u}rzburg, 97074 W{\"u}rzburg, Germany}

\author{Arianna~Poli} 
\affiliation{Dipartimento di Scienze Fisiche e Chimiche, Università dell'Aquila, Coppito-L'Aquila, Italy}

\author{Lorenzo~Crippa}
\affiliation{Institut f{\"u}r Theoretische Physik und Astrophysik and W{\"u}rzburg-Dresden Cluster of Excellence ctd.qmat, Universit{\"a}t W{\"u}rzburg, 97074 W{\"u}rzburg, Germany}
\affiliation{I. Institute of Theoretical Physics, Universität Hamburg, Notkestraße 9-11, 22607 Hamburg, Germany}

\author{Dumitru~C\u{a}lug\u{a}ru}
\affiliation{Rudolf Peierls Centre for Theoretical Physics, University of Oxford, Oxford OX1 3PU, United Kingdom}

\author{Sergio~Ciuchi}
\affiliation{Dipartimento di Scienze Fisiche e Chimiche, Università dell'Aquila, Coppito-L'Aquila, Italy}

\author{Matthias~Vojta}
\affiliation{Institut für Theoretische Physik and W{\"u}rzburg-Dresden Cluster of Excellence ctd.qmat, Technische Universit\"at Dresden, 01062 Dresden, Germany}

\author{Alessandro~Toschi}
\affiliation{Institute of Solid State Physics, TU Wien, 1040 Vienna, Austria}

\author{Giorgio~Sangiovanni}
\affiliation{Institut f{\"u}r Theoretische Physik und Astrophysik and W{\"u}rzburg-Dresden Cluster of Excellence ctd.qmat, Universit{\"a}t W{\"u}rzburg, 97074 W{\"u}rzburg, Germany}

\date{\today}

\begin{abstract}
The Kondo screening of a localized magnetic moment crucially depends on the spectral properties of the electronic bath to which it is coupled. Unlike textbook examples, realistic systems as well as dynamical mean-field theory of correlated lattice models force us to consider sharp features in the hybridization function near the Fermi energy. Divergencies of this kind can play a relevant role in twisted bilayer graphene, for which local-moment formation and isospin entropy at finite temperature are currently under the spotlight. We clarify how a low-frequency singularity impacts the screening mechanisms by means of a toy model with a tunable $\delta$-peak in the hybridization function, superimposed to a regular part.  
Our analysis unveils an unexpectedly big impact on the local-moment physics already for a parametrically small weight of the flat band in the bath.
\end{abstract}

\maketitle

\section{Introduction}
The Kondo effect is a low-temperature many-body phenomenon that arises in the presence of time-reversal symmetry when a quantum ``impurity'' spin is in contact with a metallic background of free electrons \cite{Kondo_original,Coleman_2015}. An antiferromagnetic exchange coupling between the itinerant electrons and the impurity degrees of freedom leads to the formation of singlet states, screening the localized spin below the Kondo temperature $T_{\text K}$. This many-body state influences transport not only in metals with local moments, but also in synthetic nanostructures where the Coulomb blockade in quantum dots enables the control of magnetic moments and their screening \cite{Kouwenhoven_2001, Michael_Pustilnik_2004, Pustilnik_nano, doi:10.1143/JPSJ.74.80}.

\begin{figure}[ht]
    \centering
    \includegraphics[width=\columnwidth]{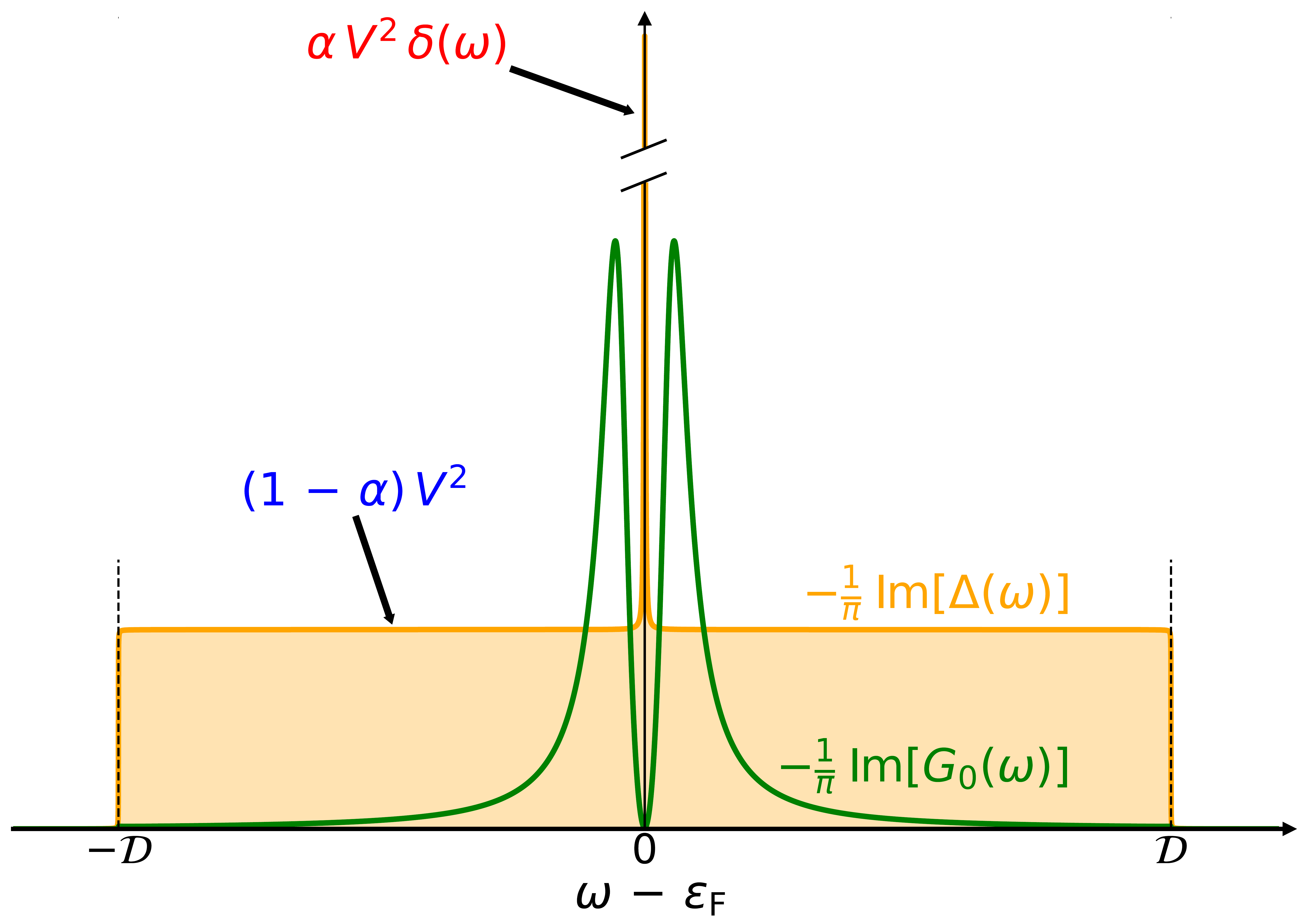}
    \caption{\justifying Toy-model hybridization function (orange) of an AIM,  Eq.~\eqref{eq:Delta_toy_model}, and corresponding noninteracting spectral function at the impurity site (green). The parameter $\alpha\,\in\,\left[0:1\right]$ tunes the weight of the $\delta$-like feature at the Fermi energy interpolating between a ``boxlike'' hybridization function ($\alpha$=0) and an isolated flat band in the bath ($\alpha$=1).}
    \label{fig:sketch}
\end{figure}

Microscopically, a general approach to Kondo physics is in terms of the Anderson impurity Hamiltonian \cite{PhysRev.124.41}. This describes both the itinerant bath and the localized degrees of freedom as electrons and couples them via the hybridization function $\Delta(\omega)$, a complex frequency-dependent quantity. In order to facilitate analytical studies of the Anderson impurity model (AIM), frequently made assumptions are (i) the infinite-bandwidth limit for the bath, i.e., $\mathcal{D}$$\rightarrow$$\infty$ in Fig.~\ref{fig:sketch}; (ii) strongly suppressed charge fluctuations on the impurity, such that the impurity behaves as a local spin; and (iii) the single-orbital nature of the impurity, implying a spin $S=1/2$.
Even though in numerical studies some of these assumptions are often relaxed, they in fact still guide our basic understanding of the Kondo effect \cite{PhysRevLett.95.066402, PhysRevResearch.3.043113, PhysRevResearch.3.043173, PhysRevLett.69.168, PhysRevLett.131.166501, Amaricci_2017, PhysRevLett.103.147205, PhysRevLett.101.166405, RevModPhys.40.380, 10.1063/1.1709517, PhysRevLett.69.1236, PhysRevLett.69.1240, PhysRevB.72.035122, PhysRevLett.97.076405, PhysRevLett.99.196403}. 

In contrast, the functional form of $-\mathrm{Im} \Delta(\omega) /\pi$ is traditionally less versatile, especially within model studies of the Kondo effect. The hybridization function close to the Fermi level $\varepsilon_\text{F}$ is either assumed to be smooth and slowly varying (a perfect constant would correspond to $\alpha$=0 in Fig.~\ref{fig:sketch}), or is taken of pseudogap form, vanishing as a power law in $(\omega-\varepsilon_\text{F})$ \cite{withoff90,PhysRevB.70.214427,lauPRX15}. Yet, the recent surge of interest in the moir\'e physics of twisted bilayer van der Waals materials \cite{SUA10, LOP07, doi:10.1073/pnas.1108174108, REN21, CHO23, PhysRevLett.131.166501} calls for an urgent reconsideration of the impact of singularities in $-\mathrm{Im} \Delta(\omega) /\pi$. 

In spite of the complexity of these systems, whose unit cells can contain up to tens of thousands of lattice sites, rather simple descriptions in terms of local moments and screening electrons are possible \cite{CAR19, CAR19a, HAU19, BER21, PhysRevB.103.205413, BER21b, doi:10.1073/pnas.1108174108, SHI22a, PhysRevLett.129.047601, C_lug_ru_2023, PhysRevB.109.045419, CHO23, HER24c, CAL25a, lauPRX15}. For twisted bilayer graphene (TBG), starting from the Bistritzer-MacDonald (BM) model on the continuum \cite{doi:10.1073/pnas.1108174108}, the so-called topological heavy-fermion (THF) periodic Anderson representation has been put forward \cite{PhysRevLett.129.047601, PhysRevLett.131.166501, C_lug_ru_2023, PhysRevX.14.031045, PhysRevB.109.045419}. A way to solve this model is through a mapping onto a single-impurity Anderson model with a self-consistently determined bath, as prescribed by dynamical mean-field theory (DMFT). 
Interestingly, the corresponding hybridization function entering DMFT calculations is characterized by a sharp low-energy feature superimposed to a (pseudogaplike) continuum -- for further details see Sec.\,\ref{sec:Bistritzer_MacDonald}. 
This is reminiscent of what happens in Kondo lattice models \cite{pruschke2000,georgesRMP,bullaRMP,youn2024hundnesstwistedbilayergraphene} and calls for a reconsideration of the impact of such peculiar form of the DMFT hybridization function on the scaling properties and on the many-body physics of TBG.

\section{Singular hybridization model}
In this section, we start from a hybridization function combining a singular and a regular contribution, without considering any DMFT self-consistency. For this reason, we analyze a single-impurity Anderson toy model with the following Hamiltonian:
\begin{equation}
\begin{split}
       \hat{H}\,&=\,\sum_{k,\sigma}\epsilon_{k}\hat{c}^{\dagger}_{k\sigma}\hat{c}_{k\sigma}+\sum_{\sigma}\epsilon_{f}\hat{f}^{\dagger}_{\sigma}\,\hat{f}_{\sigma}+U\,\hat{n}_{f\uparrow}\hat{n}_{f\downarrow}+
       \\&+\sum_{k,\sigma}\Big[V\left(\sqrt{1-\alpha}\,\hat{c}^{\dagger}_{k\sigma}\hat{f}_{\sigma}+\sqrt{\alpha}\,\hat{d}^{\dagger}_{\sigma}\,\hat{f}_{\sigma}+\text{H.c.}\right) \Big],
\end{split}
\label{eq:general_AIM_Hamiltonian}
\end{equation} 
in which the $f$-fermions describe the interacting impurity while $c_k$ and $d$ refer to a dispersive bath and to a perfectly localized level, respectively. The corresponding hybridization spectral function 
$-\mathrm{Im} \Delta(\omega) /\pi$ (orange line in Fig.~\ref{fig:sketch}) reads
\begin{equation}
    -\frac{1}{\pi}\,\text{Im}\Delta(\omega)\,=\, \left(1- \alpha\right)\frac{V^2}{2\mathcal{D}} \,\text{rect}\left(\frac{\omega}{2\,\mathcal{D}}\right) +\alpha V^2\, \delta(\omega)\,.
    \label{eq:Delta_toy_model}
\end{equation}
The parameter $\alpha$ controls the relative weight between the ``boxlike'' regular part with energy cutoff $\mathcal{D}$ and the $\delta$ function at $\varepsilon_\text{F} = 0$ (here often referred to as the ``flat band'' of the bath). $V^2$ is the overall strength of the hybridization between the impurity and the bath.
Interpreting henceforth $\omega$ --- unless differently specified --- as a complex frequency, the relation between $\Delta(\omega)$ and the noninteracting impurity Green's function $G_0(\omega)$ can be written as $\Delta(\omega)= \omega - \epsilon_f - G^{-1}_0(\omega)$. $\Delta(\omega)$ displays therefore a singularity at zero frequency if $G_0(\omega)$ vanishes for $\omega \! \rightarrow \!0$. This is captured by the AIM with the hybridization function of Eq.~(\ref{eq:Delta_toy_model}) \cite{logdiv, PhysRevLett.117.037202}. 
In the following, we solve such a toy model and draw general conclusions on how the flat band in the bath affects local-moment formation.

The $d$ level in Eq.~\eqref{eq:general_AIM_Hamiltonian} can be interpreted as representing a second, noninteracting, impurity. Therefore, the Anderson model with a $\delta$-function part of the hybridization equation \eqref{eq:general_AIM_Hamiltonian} is equivalent to a two-impurity model with a single screening channel \cite{PhysRevB.65.140405,PhysRevB.71.075305,PhysRevB.81.115316,PhysRevB.93.085428}: A concrete realization is in terms of side-coupled quantum dots where one dot is hybridized with a regular bath and coupled to a second, otherwise isolated dot where interactions are negligible. Such models have been studied theoretically in the past \cite{PhysRevB.71.075305,PhysRevB.81.115316,PhysRevB.93.085428} and have been found to display a form of two-stage Kondo screening, where the degrees of freedom on the two dots are screened at different characteristic temperatures. In the course of the paper we will relate our results to the phenomenology of such two-stage screening.

Beyond the above-mentioned case of the THF periodic Anderson model --- which will be discussed in more detail in Sec.\,\ref{sec:Bistritzer_MacDonald} --- 
flat bands in the bath of AIM, signaled by a singularity in the hybridization spectral function, are neither uncommon nor do they necessarily originate from a particular fine tuning. Rather narrow spikes in $-\mathrm{Im} \Delta(\omega) /\pi$ close to $\varepsilon_\text{F}$ generically appear in molecular systems, such as transition-metal phthalocyanines adsorbed on Ag(001) surfaces. These exhibit in the case of Mn a fairly sharp peak in the hybridization for the $xz$/$yz$ orbitals, superimposed to the more constant $z^2$ contribution \cite{NanoLettBode}.
We note that Kondo-type models with unconventional frequency behavior in the hybridization function 
have been investigated before \cite{withoff90,PhysRevB.70.214427,vojtabulla02,mitchell13}, including the case of a divergent, i.e., strongly peaked, density of states of the bath at the Fermi level \cite{vojtabulla02,mitchell13}.
However, the focus of these studies was on pure power-law behavior of the hybridization function, with no discussion on its influence on regular Kondo screening.
Kondo-insulating phases of periodic Anderson models are further examples exhibiting flat bands in the bath \cite{pruschke2000} and the same happens in DMFT calculations of three-dimensional Dirac semimetals \cite{Niklas_PRL,poliPRB109,muPRB109}. Importantly, despite the fact that, within DMFT, the hybridization function is self-consistently adjusted to enforce the mapping of the given lattice system \cite{georgesRMP}, these calculations have confirmed the robustness of the form qualitatively assumed in our toy model, up to the semimetal-to-insulator Mott transition. In addition to the above-mentioned models, further examples can be found in real materials such as Dirac and Weyl semimetals of the type of Cd$_3$As$_2$ \cite{Liang_2014}, MoP \cite{MoP} and WP$_2$ \cite{Kumar_2017} in which the protected degeneracies are located close to the Fermi level.

\section{Spin and charge susceptibilities} 
\begin{figure}
    \centering
    \includegraphics[width=\columnwidth]{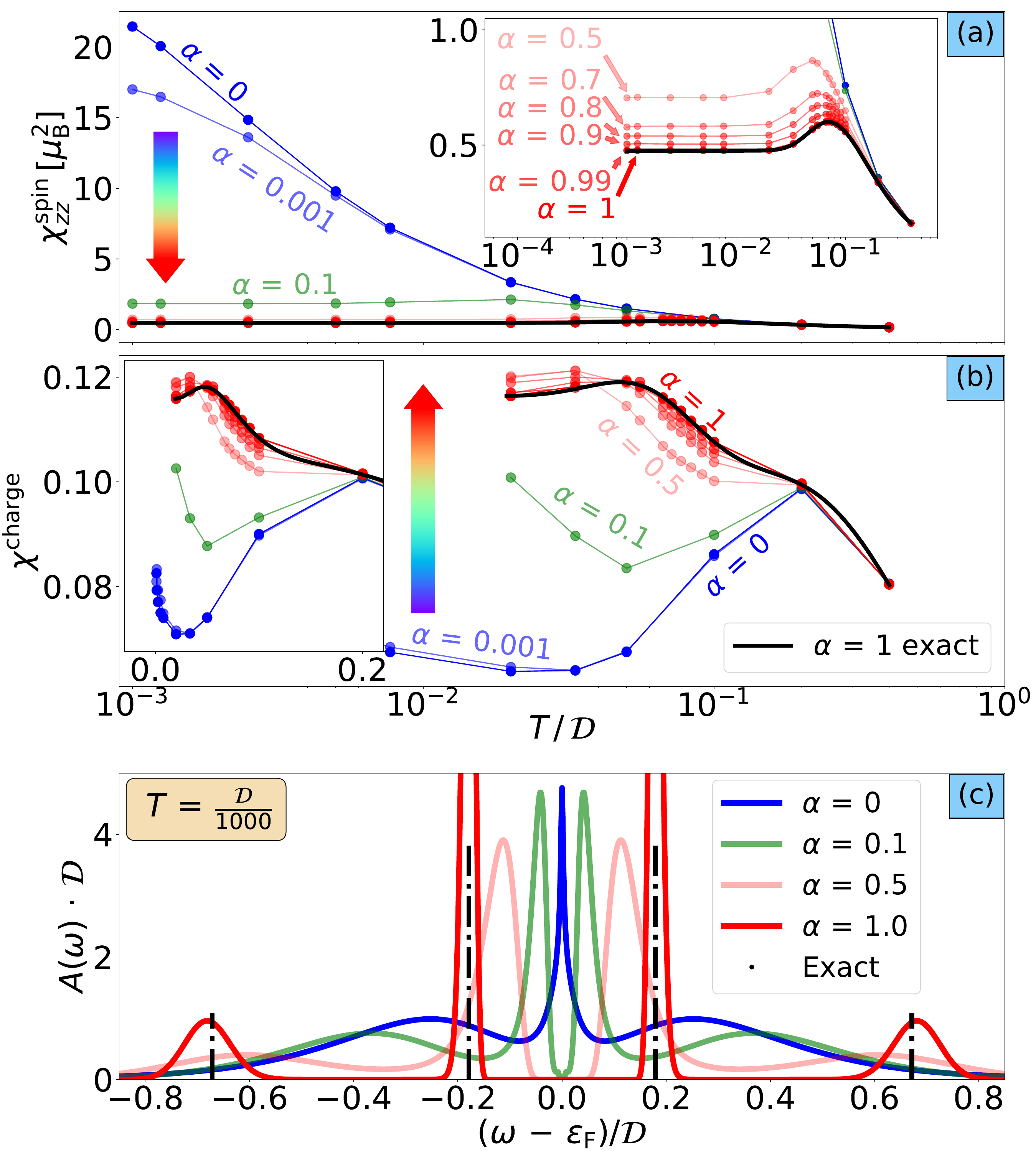}
    \caption{\justifying Local static spin $\chi^{\text{spin}}_{zz}$ (a) and charge $\chi^{\text{charge}}$ (b) susceptibility for different $\alpha$. Corresponding impurity spectra at $T$=$\mathcal{D}$/1000 as a function of the real-axis frequency, here referred to as $\omega$ (c). Other parameters: $V$=0.2$\mathcal{D}$ and $U$=0.575$\mathcal{D}$.}
    \label{fig:chi_spin_and_charge}
\end{figure}
In this section we discuss the solution of our AIM equation\,\eqref{eq:Delta_toy_model} by means of the numerically exact continuous-time quantum Monte Carlo (CTQMC) method \cite{gullRMP,Wallerberger_2019}. Later we will compare these results in some limiting cases with analytical solutions as well as against renormalization group.

A convenient way of monitoring Kondo physics is to calculate the spin-spin and charge-charge correlation functions on the impurity. 
For defining the spin correlation function, we first introduce the $S_z$-$S_z$ susceptibility (in units of $\mu_B^2$):
\begin{equation}
    \chi_{zz}^{\text{spin}}(T)\,=\,g^2\,\int^\beta_0\,\text{d}\,\tau\,\big<\hat{S}_z(\tau)\,\hat{S}_z(0)\big>\,,
\label{eq:chi_spin_impurity_tau_integration}
\end{equation}
in which $\tau$ represents the imaginary time, ${\hat{S}_z(\tau)\,=\,\frac{1}{2}\,\left(\hat{n}_{\uparrow}(\tau)\,-\,\hat{n}_{\downarrow}(\tau)\right)}$ the $z$ component of the spin operator and $\hat{n}(\tau)$ the impurity occupation number operator. Further, $g$ is equal to 2 and $\beta\,=\,1/T$ is the inverse temperature. The charge susceptibility reads instead:
\begin{equation}
\begin{split}
        \chi^{\text{charge}}(T)=&\int_{0}^\beta\,\text{d}\tau\,\left<\left[\hat{n}_{\uparrow}(\tau)+\hat{n}_{\downarrow}(\tau)\right]\left[\hat{n}_{\uparrow}(0)+\hat{n}_{\downarrow}(0)\right]\right>\\&-\,\beta\,\left<\left(\hat{n}_{\uparrow}+\hat{n}_{\downarrow}\right)\right>^2\,.
\end{split}
\end{equation}
Both quantities are calculated with CT-QMC for various values of $\alpha$.
\begin{figure}[tb!]
    \centering
    \includegraphics[width=\columnwidth]{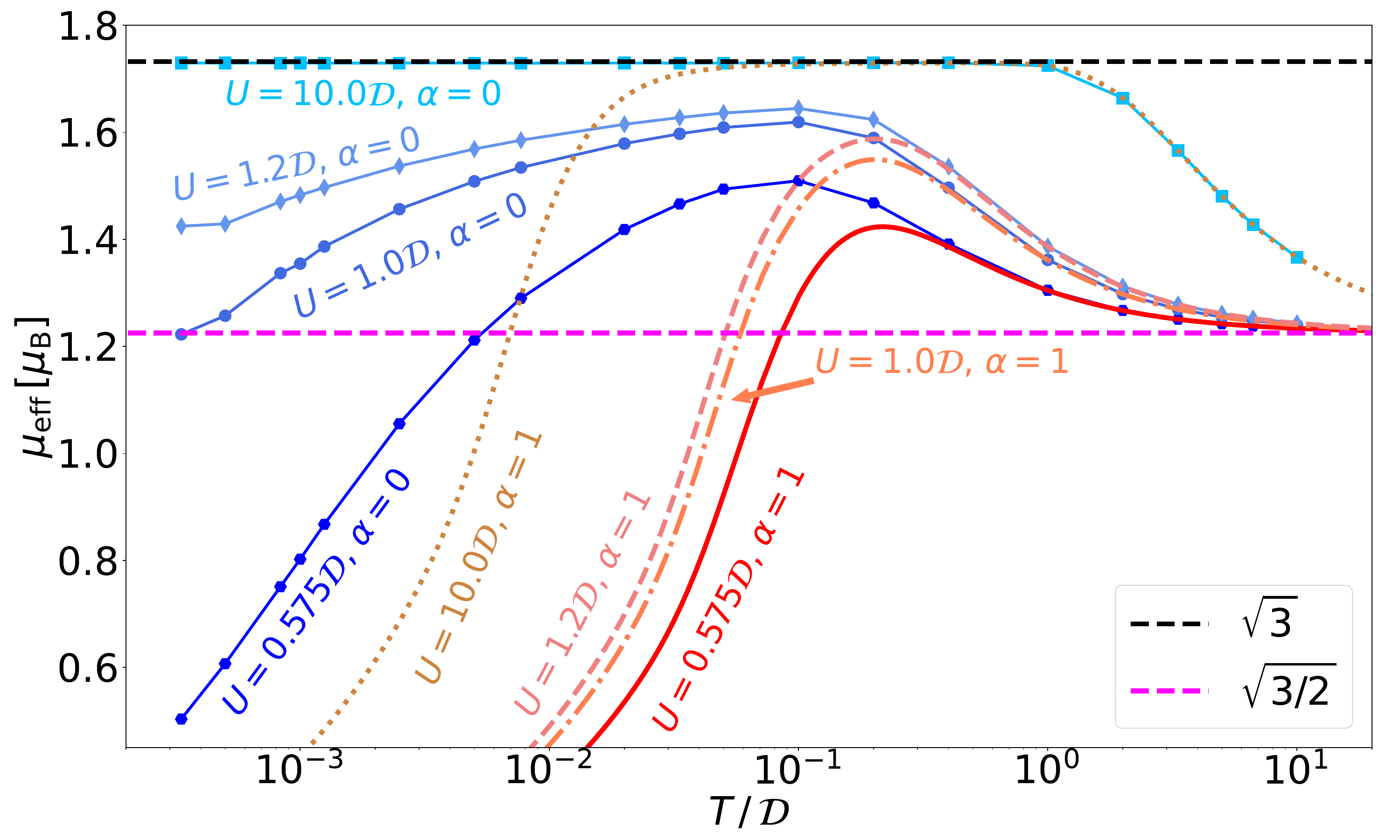}
    \caption{\justifying Effective moment $\mu_{\text{eff}}$ of the impurity for different values of $U$ from $\alpha$=0 (blue) to $\alpha$=1 (red).} 
    \label{fig:mu_eff_different_U}
\end{figure}

We start by showing the static spin- and charge-susceptibilities $\chi^{\rm spin}_{zz}(T)$ and $\chi^{\rm charge}(T)$ at finite temperature $T$ for different values of $\alpha$ and for the strength of the repulsive interaction $U$. The average occupation of the impurity is fixed to 1.

At intermediate-to-high temperatures, a $1/T$ Curie behavior for the spin is expected, characteristic of unscreened local magnetic moments. Upon lowering $T$, a crossover to a constant Pauli-like $\chi^{\rm spin}_{zz}(T)$ signals the onset of screening \cite{RevModPhys.47.773,PhysRevB.21.1003,Bulla_1997}. Charge fluctuations are suppressed when the local moment is formed and display a minimum around $T_\text{K}$. Upon further lowering the temperature, a ``revival'' in $\chi^{\rm charge}$ appears, as counterpart of the Pauli plateau in the spin \cite{PhysRevLett.126.056403, PhysRevB.105.L081111, PhysRevB.101.165105, 10.21468/SciPostPhys.16.2.054}.
In Fig.~\ref{fig:chi_spin_and_charge}, this behavior of the two quantities is recognizable in the case of the conventional boxlike hybridization ($\alpha$=0). The deviation from the Curie law is clearly visible \cite{QMC, kuglerPRL124, grunderPRB112} and the minimum in the charge susceptibility is instead nicely resolved. A value of $\alpha$ of only 5\%--10\% induces a huge effect on the spin and charge responses with respect to $\alpha$=0: $\chi^{\rm spin}_{zz}(T)$ displays an extended plateau which is reached at temperatures more than two orders of magnitude higher than the onset of Pauli behavior for $\alpha$=0. At the same time, charge fluctuations are much less suppressed and display no minimum with the ``revival'' upturn, as for $\alpha$=0. A $\delta$-like peak in the  hybridization spectral function $-\mathrm{Im} \Delta(\omega) /\pi$ causes therefore an unexpectedly rapid and strong qualitative deviation from the Curie behavior.

The intrinsic fragility of the local moment also manifests as a splitting of the Kondo peak which, as shown in Fig.\,\ref{fig:chi_spin_and_charge}(c), yields a four-peak structure already for the smallest values of $\alpha$ at which the susceptibility changes qualitatively. 
Note that here the $\alpha$=1-spectrum is shown both as ``poles and weights'' from the exact solution as well as through CTQMC followed by maximum-entropy analytic continuation \cite{jarrellMaxEnt,kaufmannAnaCont}, which yields an intrinsic broadening. The perfect matching between the two cases nicely validates our analytic continuation procedure also for $\alpha$$\neq$1.

Let us stress here, that the splitting of the Kondo peak can also be viewed as a consequence of the change of the scattering phase shift from $\pi/2$ to $\pi$ due to the $\delta$-function in the hybridization \cite{Hewson_1993}, similar to what happens in two-impurity models \cite{PhysRevB.81.115316}. This is a robust feature that persists also within DMFT calculations of lattice models exhibiting a flat band in the bath \cite{Niklas_PhD}. Yet, for those lattice models, the DMFT self-energy also adds a finite-temperature broadening to the low-energy peak. This smoothens the phase-shift jump and makes the splitting of the Kondo peak strongly dependent on the ratio between temperature and interaction strength (see also Appendix \ref{App:Scattering_phase_shift}). 

\begin{figure}[tb!]
    \centering
    \includegraphics[width=\columnwidth]{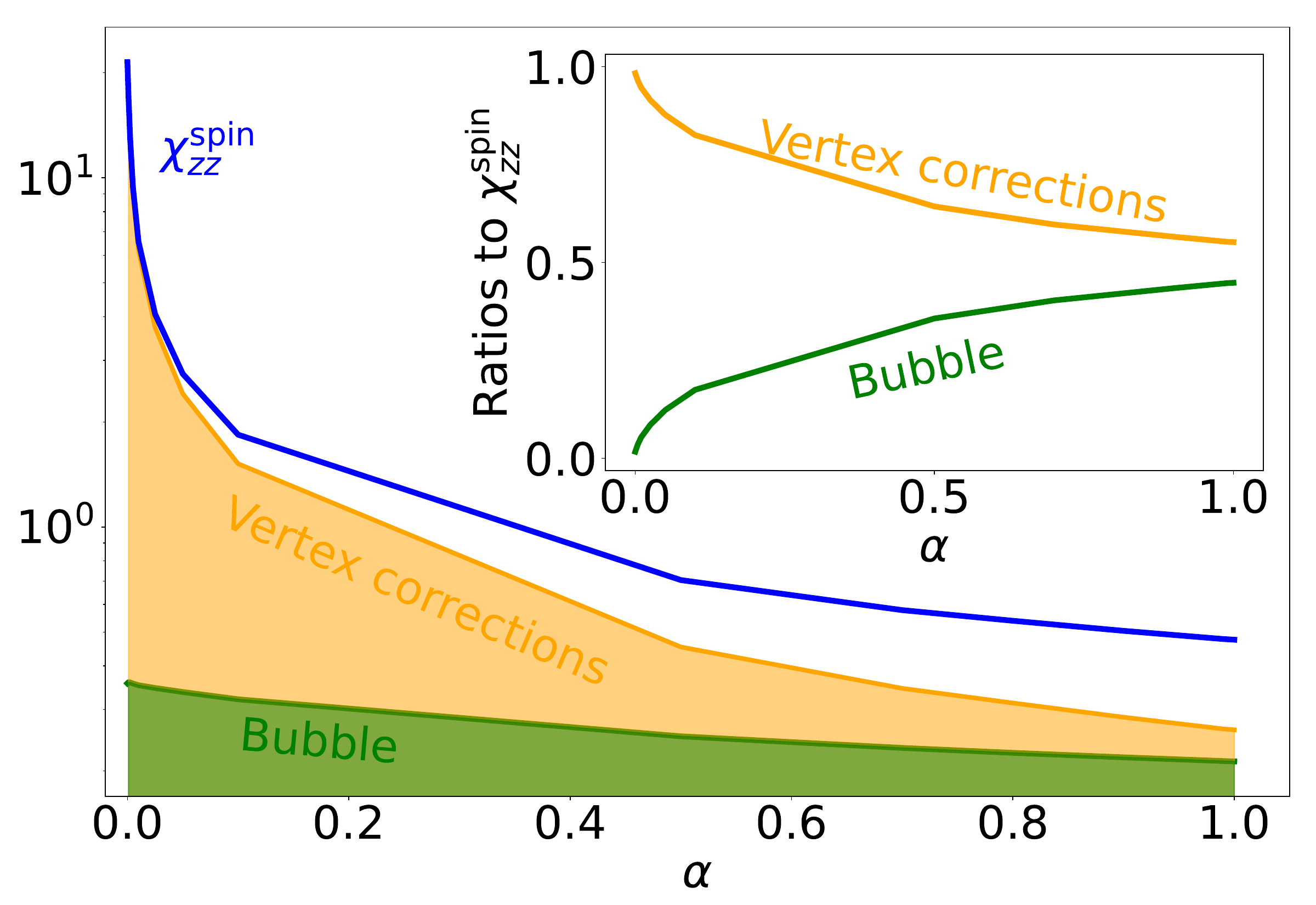}
    \caption{\justifying Local spin susceptibility $\chi^{\text{spin}}_{zz}$ (blue) together with the corresponding contributions of bubble (green) and vertex corrections (orange) at ${T\,=\,0.001\,\mathcal{D}}$ and ${U\,=\,0.575\,\mathcal{D}}$. Inset: The ratios corresponding to $\chi^{\text{spin}}_{zz}$ show how the two contributions have a qualitatively different dependence on $\alpha$.}
    \label{fig:vertex_corrections}
\end{figure}
\section{Effective moment} 
The impact of the singularity in $\Delta(\omega)$ on the local-moment formation can be appreciated by looking at the respective temperature dependence at different values of $U$. Figure~\ref{fig:mu_eff_different_U} shows this for $\alpha$=0 and 1 in terms of the local magnetic moment of the impurity $\mu_{\text{eff}}\,=\,g\,\mu_{\text{B}}\,\sqrt{S\,(S\,+\,1)}$, where $S$ denotes the effective spin on the interacting site. The spin susceptibility $\chi_{zz}^{\text{spin}}(T)$ goes at high temperatures as $\mu^2_{\text{eff}}/{(3T)}$.
If we start from the ``infinite''-$T$ region of the interaction-scale $U$, all curves reach the ``free'' moment $\mu_{\text{eff}}\,=\,\sqrt{3/2}\,\mu_{\text{B}}$, as illustrated in Fig.~\ref{fig:mu_eff_different_U}. This corresponds to the many-body configurations with the different number of electrons being only differentiated by the thermal distribution. Reducing $T$ brings $\mu_{\text{eff}}$ up to the nominal large-$U$ Curie-Weiss value, which for a half-filled $S$=$1/2$ impurity is $\sqrt{3} \, \mu_{\text{B}}$, as we are considering only the $z$-component of the spin susceptibility. If $U/\mathcal{D}$ is not huge, this value is not fully reached because of the still active charge fluctuations in the AIM.  
Eventually, upon further decreasing $T$, the screening of the moment takes place. As shown in Fig.~\ref{fig:mu_eff_different_U}, this process happens very differently between $\alpha$=0 and 1: in the conventional $\alpha$=0 case, the plateau is left only when $T$ starts to approach the $T_\text{K}$ from above, i.e. at temperatures that for these values of $U$ and $\mathcal{D}$ are below 10$^{-3}\mathcal{D}$, consistent with the universal Kondo temperature curve \cite{PhysRevB.21.1003, Hewson_1993}. On the contrary, for $\alpha$=1, i.e. in the extreme case of a bath made of the $\delta$-function only, the loss of local moment happens at temperatures two orders of magnitude higher, with a much weaker dependence on $U$ and with a much higher slope w.r.t. $T$. 

\section{Vertex corrections and bubble contribution}
Physically, the loss of local moment for finite values of $\alpha$ can be understood as follows: A $\delta$-like hybridization corresponds to a single isolated flat band in the bath. If this sits at $\varepsilon_{\text F}$ as in the case of TBG at charge neutrality, this level hosts one electron which has a strong tendency to form a singlet with the impurity spin \cite{Hewson_1993}. Increasing the relative weight of the hybridization peak therefore reduces the many-body nature of the whole system, and hence the fluctuating local moment, in a very efficient manner. This can be illustrated by quantifying the impact of $\alpha$ on the bubble of two dressed impurity Green's functions as well as on the vertex corrections. These correspond to diagrams in which the two dressed Green's functions are connected by interaction lines, i.e. processes that cannot be described just as a bubble of two Green's functions. In Fig.~\ref{fig:vertex_corrections} one sees clearly how the vertex corrections, which are almost the only diagrams responsible for the local moment at $\alpha$=0 (here taking the example of $T=0.001\mathcal{D}$) drop already by one order of magnitude at $\alpha$=0.1. At the same time the bubble contribution decreases with $\alpha$ very gradually instead, so that its relative importance effectively increases.

\vspace{-0.4cm}
\section{Perturbative renormalization group}
\label{sec:RG}
To substantiate the above discussion, we apply the perturbative renormalization group (RG). This allows us to quantify the relative importance of the hybridization to the $\delta$ function with respect to the conventional boxlike bath in driving the system toward the Kondo fixed point. For finding the flow equations we use a Kondo Hamiltonian with two types of fermions, derived from the above Anderson model,
\begin{equation}
\begin{split}
    \hat{H}=&\sum_{k,\sigma}\epsilon_k\hat{c}^\dagger_{k\sigma}\hat{c}_{k\sigma}+\sum_{\sigma,\sigma'} \Big\{ \mathbf{S}\cdot\frac{\boldsymbol{\tau}_{\sigma,\sigma'}}{2} \\&\cdot \Big[ J\hat{c}^\dagger_{0\sigma}\hat{c}_{0\sigma'}+K\hat{d}^\dagger_{\sigma}\hat{d}_{\sigma'}+M(\hat{c}^\dagger_{0\sigma}\hat{d}_{\sigma'}+\text{H.c.})\Big] \Big\}\,,
\end{split}
\label{eq:Kondo_Hamiltonian}
\end{equation}
with $\mathbf{S}$ being the impurity spin and $\boldsymbol{\tau}$ the vector of Pauli matrices. The couplings $J \!\!\propto\!\!(1-\alpha)$, $K\!\!\propto\!\!\alpha$ and $M \!\!\propto \!\!\sqrt{\alpha(1-\alpha)}$ are associated respectively with the constant hybridization, with the $\delta$ peak and with the corresponding mixed term. $\hat{c}^{(\dagger)}$ and $\hat{d}^{(\dagger)}$ are the annihilation (creation) operators for the two baths while $\hat{c}^{(\dagger)}_{0\sigma}$ indicates the Fourier transformed operators $\hat{c}^{(\dagger)}_{k\sigma}$ at the impurity site.
\begin{figure}[t]
    \centering
    \includegraphics[width=\columnwidth]{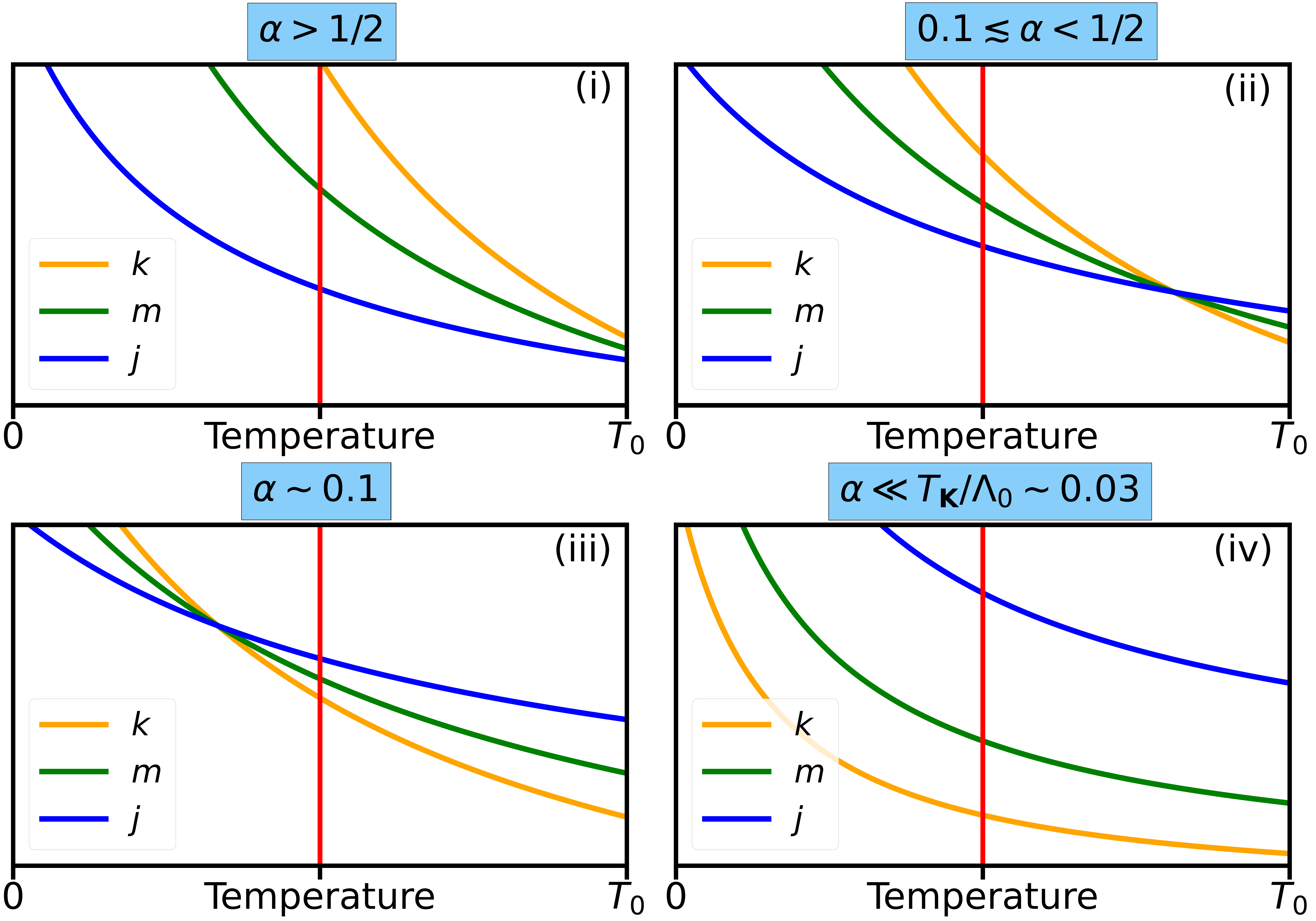}
    \caption{\justifying Sketch of the RG flow of the dimensionless couplings $k$, $j$ and $m$ to the $\delta$ peak, the broad band and the hybridization between both, respectively, showing four regimes of the $\delta$ weight $\alpha$.
    (i) For large $\alpha>1/2$ the $\delta$-peak coupling $k$ dominates at any $T$. (ii,iii) At smaller $\alpha$ the couplings cross as a function of $T$, implying a crossover which is located above or below the relevant temperature scale for the Kondo effect (red line).
    (iv) For very small $\alpha\ll T_\text{K}/\Lambda_0$ the $j$ coupling dominates at any $T$ where $T_\text{K}$ describes the Kondo temperature of the regular hybridization. The $m$ coupling is always in between the other couplings.
    }
    \label{fig:Sketch_RG_alpha_temperature_regimes}
\end{figure}
Using this model, the RG flow of the dimensionless couplings ${j\,=\,J/\Lambda_0}$, ${k\,=\,K/\Lambda_0}$ and ${m\,=\,M/\Lambda_0}$ yields the following $\beta$-functions:
\begin{equation}
\begin{split}
    \beta(j)\,& \coloneq \,\Lambda\,\frac{\text{d}j}{\text{d}\Lambda}\,=\,-j^2,\\
    \beta(k)\,&\coloneq \,\Lambda\,\frac{\text{d}k}{\text{d}\Lambda}\,=\,-k - m^2,\\
    \beta(m)\,&\coloneq \,\Lambda\,\frac{\text{d}m}{\text{d}\Lambda}\,=\,-\frac{m}{2} - jm.
\end{split}
\label{eq:flow_equations}
\end{equation}
These equations can be integrated numerically returning the running couplings $j,k,m(\Lambda)$. In order to draw conclusions on the dominant couplings, we identify the running cutoff $\Lambda$ with the temperature $T$. Using the flow equations, we can hence determine the temperature scale at which conventional Kondo screening dominates over singlet formation.
Given that $k$ grows faster upon cooling than the other couplings, an initially small $k$ can become the dominant coupling at low temperature. This leads to four different regimes for $\alpha$, as shown in Fig.~\ref{fig:Sketch_RG_alpha_temperature_regimes}, with regimes (iii) and (iv) corresponding to two-stage screening in the context of two-impurity models \cite{PhysRevB.71.075305,PhysRevB.81.115316,PhysRevB.93.085428}:

\begin{itemize}
\item
(i) For $\alpha>1/2$ the flat-band coupling $k$ dominates for all $T$, implying that conventional Kondo screening is absent.

\item
(ii,iii) For intermediate $\alpha$ we see that $j$ dominates at elevated $T$, but $k$ wins at low $T$. Therefore, conventional Kondo screening is present, but cut off by the flat-band effects at a small energy or temperature scale, leading e.g. to a splitting of the Kondo peak. Regimes (ii) and (iii) are distinguished by whether this crossover happens above or below $T_\text{K}=\Lambda_0 e^{-\frac{1}{j_0}}$, the Kondo temperature associated with the regular part of the hybridization (see red vertical lines in Fig.~\ref{fig:Sketch_RG_alpha_temperature_regimes}).

\item
(iv) For very small $\alpha$ conventional Kondo screening dominates down to lowest $T$; i.e., $j$ flows to strong coupling before $k$ and $m$ become sizable. If we ignore the cross-coupling terms in the RG equations, we can estimate regime (iv) as $\alpha \ll T_K/\Lambda_0$. We note that finite nonzero $\alpha$ will still lead to a small splitting of the Kondo peak.

\end{itemize}

The four different regimes can be also directly related to the behavior of the spin and charge susceptibilities in our numerical results in Fig.\,\ref{fig:chi_spin_and_charge}. Cases (i) and (iv) are connected to situations with $\alpha\lesssim 0.03$ and $\alpha>1/2$, yielding the standard Kondo effect or the immediate screening of the moment by direct singlet formation for all $T$, respectively. The change from regime (ii) to (iii) occurs around $\alpha\sim 0.1$. This result supports the behavior of the susceptibilities in Fig.\,\ref{fig:chi_spin_and_charge} for extremely small $\alpha$ values where the formation of local moments is still present.

\section{Hybridization function for the Bistritzer-MacDonald model}
\label{sec:Bistritzer_MacDonald}
In this section we derive the hybridization function entering DMFT calculations of TBG \cite{PhysRevLett.131.166501,bascones,youn2024hundnesstwistedbilayergraphene,PhysRevX.14.031045}. In particular we will examine the low-energy sharp feature originating from the mapping of the BM model onto a single impurity, similar to what happens for Kondo insulators \cite{pruschke2000,youn2024hundnesstwistedbilayergraphene}. Let us recall that we will not be considering the feedback of the DMFT self-consistency here. The reason is that we just want to focus on the influence of the low-frequency structure of $\Delta(\omega)$ on the screening of the local moment, as analyzed in the previous sections based on our toy-AIM. For this reason, we will inspect the form of $-$Im$\Delta(\omega)/\pi$ of the BM model focusing at low frequencies and for different twist angles all the way down to the magic angle.

Before proceeding, it is also important to note that we are not interested here in determining whether or not Kondo-insulating solutions can appear in self-consistent DMFT solutions of the THF model depending on filling factor $\nu$, temperature and other relevant parameters for TBG. 
Similarly, the multi-orbital and multi-valley nature of THF as well as the complexity of the interaction terms are not taken into account by our toy model, not to mention the effect of symmetry breaking and strain \cite{CrippaStrain}. 
Further, at $\nu=0$ the tendency toward a Mott solution is strong and therefore the $\delta$-like contribution in Im$\Delta(\omega)$ may be eventually weakened or completely washed out by self-consistency.

The purpose of this study is instead to point out how even a residual small contribution of the singular part (corresponding to our dimensionless parameter $\alpha$ as small as 0.1) can have a significant effect on the size and nature of the local moment. In the THF model, it is known that the local moments are small \cite{PhysRevX.14.031045} primarily because two electrons pay the same Coulomb energy by occupying the $f$-like flat bands, irrespective of their orbital and their spin flavor. As a consequence, many-electron high-spin configurations are not favored ($J_{\rm Hund} \!= \!0$).
However, based on our findings, the $\delta$-like part of the hybridization, even if strongly suppressed by the DMFT self-consistency, may further contribute to weaken local-moment physics with the mechanism described here.

To obtain the DMFT hybridization function of the THF periodic Anderson model, we indeed start from the continuum BM model and project it onto a single-impurity Anderson model defined on the $f$-like Wannier subspace introduced by Song and Bernevig for the THF model \cite{PhysRevLett.129.047601}.
In brief, the THF model is essentially a basis transformation of the low-energy TBG spectrum, in which its active bands emerge from the hybridization between localized, heavy and strongly correlated $f$ fermions and topological, itinerant and strongly dispersive $c$ electrons. In the THF basis, the single-particle Hamiltonian of the system is given by~\cite{PhysRevLett.129.047601,C_lug_ru_2023}
\begin{widetext}
\begin{equation}
    \label{app:eqn:thf_non-int_ham}
    \hat{H}_{0}(\mathbf{k})\,=\,\sum_{\vert{\mathbf{k}}\vert{}<\Lambda_c}\,\sum_{a a' \eta s} H^{(c,\eta)}_{a a'}(\mathbf{k})\,c^{\dagger}_{\mathbf{k} a \eta s}c_{\mathbf{k} a' \eta s} + \frac{1}{\sqrt{N}} \sum_{\vert{\mathbf{k}}\vert{}<\Lambda_c}\,\sum_{\alpha a \eta s} \left[e^{i \mathbf{k}\cdot\mathbf{R} - \frac{\vert{\mathbf{k}}\vert{}^2 \lambda^2}{2}} H^{(fc,\eta)}_{\alpha a}(\mathbf{k}) f^{\dagger}_{\mathbf{R} \alpha \eta s} c_{\mathbf{k} \alpha \eta s} + \text{H.c.}\right]\,.
\end{equation}
\end{widetext}
In Eq.~(\ref{app:eqn:thf_non-int_ham}), $\cre{c}{\mathbf{k} a \eta s}$ (for $1 \leq a \leq 4$) and $\cre{f}{\mathbf{R} \alpha \eta s}$ (for $\alpha = 1,2$) denote the creation operators for the $c$ electrons at momentum $\mathbf{k}$ and the $f$ electrons at lattice site $\mathbf{R}$, respectively. The Hamiltonian is diagonal in both the spin $s=\uparrow,\downarrow$ and valley $\eta=\pm$ indices. Because they are strongly dispersive, only the $c$ electrons with small momenta are relevant for the low-energy physics of the model; hence, a momentum cutoff $\Lambda_c$ is imposed for the $c$ electrons. The $f$ electrons are dispersionless but couple to the $c$ electrons through the second term in Eq.~(\ref{app:eqn:thf_non-int_ham}). The hybridization function decays exponentially with momentum due to the finite spatial spread of the $f$ orbitals~\cite{PhysRevLett.129.047601,C_lug_ru_2023}, characterized by $\lambda$, the spread of the $f$ electron Wannier orbitals.

The corresponding matrix elements of the $c$ fermions have the form:
\begin{equation}\label{Hc}
    H^{c,\eta}_{a a'}(\mathbf{k})\!=\!\!\begin{pmatrix}
        0_{2\times2} && v_* (\!\eta k_x \sigma_0 \!+\! i k_y \sigma_z\!) \\ v_* (\!\eta k_x \sigma_0 \!-\! i k_y \sigma_z\!) && M \sigma_x 
    \end{pmatrix} 
\end{equation}
and the hybridization between $f$ and $c$ fermions reads
\begin{equation}\label{Hcf}
    H^{cf,\eta}_{a \alpha}(\mathbf{k})\,=\,\begin{pmatrix}
        \gamma \sigma_0 + v'_* (\eta k_x \sigma_x + k_y \sigma_y) \\ v''_* (\eta k_x \sigma_x - k_y \sigma_y)
    \end{pmatrix}\,.
\end{equation}
Here, all parameters are described within the context of the aforementioned papers. Remarkably, the THF mapping is applicable over a broad range of parameters, both at~\cite{PhysRevLett.129.047601} and away~\cite{C_lug_ru_2023} from the magic angle, and in the presence or absence of strain or lattice relaxation effects \cite{xv3m-vtlr}. Therefore, to compute the DMFT hybridization function for the localized $f$ electron, one can, in principle, work directly within the THF model. However, away from magic angle, a nearest-neighbor $f$ electron hopping must be included (i.e. $H^{ff}$ would no longer be zero), and a sufficiently large cutoff $\Lambda_c$ is required to ensure that the low-energy dispersion of the active bands near the Dirac points is accurately reproduced.
\begin{figure*}
    \centering
    \includegraphics[width=1.00\textwidth]{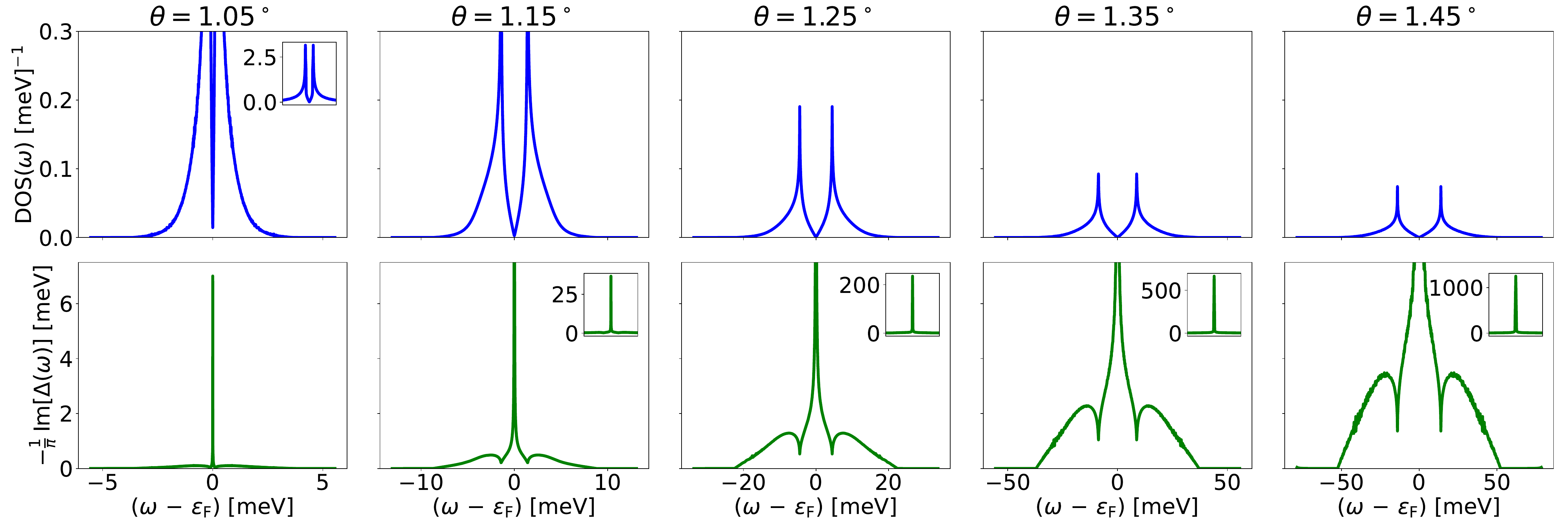}
    \caption{\justifying Impurity DOS$(\omega)$ and $-\frac{1}{\pi}\,\text{Im}[\Delta(\omega)]$ for twist angles of $\theta\,=\,1.05^{\circ}$, $1.15^{\circ}$, $1.25^{\circ}$, $1.35^{\circ}$ and $1.45^{\circ}$ (left to right) of twisted bilayer graphene for the THF model. The model always exhibits a two-dimensional linear crossing at the $K$ point. In the hybridization function the zero weight of $-\frac{1}{\pi}\,\text{Im}\left[G_0(\omega)\right]$ at the Fermi level yields a peak superimposed on a smooth background, with a twist angle dependent weight. The insets display the full $y$ range of the corresponding functions, while the frequency range for the DOS is set at [$-1$m$eV$;$+1$m$eV$] and for $-\frac{1}{\pi}\,\text{Im}[\Delta(\omega)]$ at [$-5$m$eV$;$+5$m$eV$]. The values of the small imaginary part $0^{+}$ added to the real frequencies that we used are: $0^{+}\,=\,0.001$ for magic angle ($\theta\,=\,1.05^{\circ}$) and $0^{+}\,=\,0.01$ for other twist angles.}
    \label{fig:DOS_hyb_func_magic_angle_1_40}
\end{figure*}

To circumvent these issues, we instead employ the continuum BM Hamiltonian~\cite{doi:10.1073/pnas.1108174108}, which reads
\begin{equation}
	\label{app:eqn:plane_wave_ham}
	\hat{H}_{0}=\sum_{\mathbf{k}, \eta, \alpha, \beta, s} \sum_{\mathbf{Q},\mathbf{Q}' \in \mathcal{Q}_{\pm}} \left[h^{\eta}_{\mathbf{Q},\mathbf{Q}'} \left( \mathbf{k} \right) \right]_{\alpha \beta} \cre{c}{\mathbf{k},\mathbf{Q},\eta,\alpha,s} \des{c}{\mathbf{k},\mathbf{Q}',\eta,\beta,s},
\end{equation}
where the definition of the plane-wave operators $\cre{c}{\mathbf{k},\mathbf{Q},\eta,\alpha,s}$, as well as the explicit form of the corresponding plane-wave Hamiltonian, is given in Ref.~\cite{PhysRevB.103.205413}. In the plane-wave basis, the wave function of the $f$ electrons is expressed as
\begin{equation}
	\label{app:eqn:f_fermions_mom_def}
	\cre{f}{\mathbf{k},\alpha,\eta,s} = \sum_{\mathbf{Q},\beta} v^{\eta}_{\mathbf{Q}\beta;\alpha} \left( \mathbf{k} \right) \cre{c}{\mathbf{k},\mathbf{Q},\beta,\eta,s},
\end{equation} 
with the coefficients $v^{\eta}_{\mathbf{Q}\beta;\alpha} \left( \mathbf{k} \right)$ obtained through a combined Wannierization and disentangling procedure. Using the plane-wave Hamiltonian and the wave function of the $f$ electrons, we can obtain the noninteracting $f$ electron Green's function. Specifically, we first compute the single-particle Green's function of the plane-wave Hamiltonian from Eq.\,(\ref{app:eqn:plane_wave_ham}),  
\begin{equation}\label{Geta}
	\mathcal{G}^{\eta} \left( \omega, \mathbf{k} \right) = \left[  \omega   \one - h^{\eta} \left( \mathbf{k} \right) \right]^{-1},
\end{equation}
where $\omega$ is to be understood here either as a zero-temperature continuum Matsubara frequency or as $\omega+i 0^{+}$ where $0^{+}$ is a small positive shift to obtain the retarded Green's function.
Equation~(\ref{Geta}) can be straightforwardly projected onto the $f$-electron states
\begin{equation}
	\resizebox{0.48\textwidth}{!}{$\mathcal{G}^{f,\eta}_{\alpha \beta} \left( \omega, \mathbf{k} \right) = \sum_{\mathbf{Q},\mathbf{Q}' \in \mathcal{Q}_{\pm}} v^{*\eta}_{\mathbf{Q}\alpha';\alpha} \left( \mathbf{k} \right) \left[ \mathcal{G}^{\eta}_{\mathbf{Q},\mathbf{Q}'} \left( \omega, \mathbf{k} \right) \right]_{\alpha' \beta'} v^{\eta}_{\mathbf{Q}'\beta';\beta} \left( \mathbf{k} \right).$}
\end{equation}
The $ff$ block of the on-site noninteracting Green's function, which is orbital and valley diagonal, is subsequently obtained through Fourier transformation
\begin{equation}
	\label{app:eqn:greens_func_local}
	\frac{1}{N} \sum_{\mathbf{k}} \mathcal{G}^{f,\eta}_{\alpha \beta} \left( \omega, \mathbf{k} \right) = G_{0}\!\left( \omega \right)\!{\big\vert{}}_{ff}  \delta_{\alpha \beta}.
\end{equation}
In our numerical simulations we employed an $N = 1536 \times 1536$ $\mathbf{k}$ mesh and computed the noninteracting Green's function of the $f$ fermions at half filling and particle-hole symmetry according to Eq.\,(\ref{app:eqn:greens_func_local}). The hybridization function for the $f$ fermions without feedback effects from the DMFT self-consistency is then defined as
\begin{equation}\label{hyb_fun_def}
    \Delta(\omega)\,=\,\omega -\,\left[G_0(\omega){\big\vert{}}_{ff}\right]^{-1}\,.
\end{equation}
In Fig.\,\ref{fig:DOS_hyb_func_magic_angle_1_40} we show some plots of the projected $-\text{Im}G_0(\omega)/\pi$ and of the corresponding DMFT hybridization function, choosing different twist angles. The form of $-\text{Im}\Delta(\omega)/\pi$ is hence of Withoff-Fradkin type \cite{withoff90, PhysRevB.70.214427}, characteristic of a pseudogap coming from the form of the $c$ bands \cite{lauPRX15} plus a low-energy sharp peak coming from the mapping of a periodic Anderson model with vanishing density of states of the auxiliary DMFT bath. This also holds for a full self-consistent DMFT solution of the THF model, as long as the feedback of the self-energy for a given set of parameters does not qualitatively change the low-frequency structure of the hybridization function of the auxiliary impurity problem \cite{Niklas_PRL,Niklas_PhD,PhysRevLett.95.066402}. This peak can be resolved more easily at larger twist angles, since the weight of the peak increases with the twist angle. 
Nonetheless, its impact on the local-moment formation will be strong, as demonstrated in the first part of this work.

\section{Conclusion}
Using a combination of numerics and weak-coupling RG, we have analyzed how a sharp peak in the hybridization function of an Anderson impurity model qualitatively alters the Kondo physics. The surprising result of our analysis is that the local moment, as well as its gradual loss upon cooling via Kondo screening, is completely suppressed by the singular hybridization even if the relative weight of the latter is as small as about 10\% of the total hybridization strength. We find that the effects of a flat band in the bath become unimportant only if its weight is smaller than the dimensionless Kondo temperature associated with the nonsingular part of the bath.  
This is relevant for any DMFT calculation in which the auxiliary self-consistently determined bath displays similar sharp features. In general, we note that finite temperature effects can broaden $\delta$-like peaks in the DMFT hybridization function, which could smoothen, in principle, the effects described above on the local moment. Nevertheless, the physics analyzed here is robust against the addition of such finite width, as detailed in Appendix~\ref{App:Scattering_phase_shift}. In particular, the actual destiny of the local moment at further lower temperatures will quantitatively depend on the hierarchy between the temperature, weight and width of the sharp peak.
Our results indicate a deep impact of the hybridization to such isolated bath level on the many-body nature of the Kondo effect: Vertex corrections, which are crucial in the conventional case to build up the local moment, are particularly fragile against the $\delta$-peak in the bath. Hence Kondo screening does not take place despite the still dominant weight of the nonsingular bath states.

The motivation for this study comes mainly from the physics of fluctuating local moments of TBG and the associated entropy release which plays a crucial role for the reported isospin Pomeranchuk effect \cite{Pomeranchuk, PhysRevB.48.7167, Biao_Lian, Rozen_2021, Saito_2021, PhysRevX.14.031045, CrippaStrain}.
One interesting point is why in TBG --- at fixed temperature and strain level --- such local moments are not particularly big in size. The primary reason for that is the absence of $J_{\rm Hund}$ in the Coulomb interaction within the $f\!-\!f$ sector.  
The results presented here for the Anderson impurity toy model hint at the possible relevance of an additional concurring mechanism, namely the sharp low-frequency feature that may survive the DMFT self-consistency in some regions of the phase diagram of TBG. In future studies, this aspect will be further investigated within low-temperature DMFT solutions of the THF model, with specific attention on narrow structures becoming visible increasing the resolution on the real-frequency axis. Moreover, the effect of breaking of the particle-hole symmetry and of finite doping away from charge neutrality will be interesting to address. In this regard, in addition to strain, electron-phonon interaction would add one further energy scale and hence is also expected to play a role.

\begin{acknowledgments}
We thank Sergio Caprara for useful discussions. L.C. and G.S. are grateful for important conversations and collaborations with B. Andrei Bernevig, Haoyu Hu, Roser Valent\'i, Gautam Rai, Tim Wehling, Luca de' Medici and Antoine Georges. 
M.F., L.C., M.V., and G.S. acknowledge financial support by the Deutsche Forschungsgemeinschaft (DFG, German Research Foundation) under Germany's Excellence Strategy through the W\"urzburg-Dresden Cluster of Excellence on Complexity, Topology and Dynamics in Quantum Matter-ctd.qmat (EXC 2147, project-id 390858490) and M.~V. through SFB 1143 (project-id 247310070). L.C. acknowledges support from the Cluster of Excellence CUI: Advanced Imaging of Matter (EXC 2056, project-id No. 390715994) and SPP 2244 (WE 5342/5-1 project-id No. 422707584). S.~C. acknowledges funding from NextGenerationEU National Innovation Ecosystem grant ECS00000041 - VITALITY - CUP E13C22001060006 and grant PE00000023 - IEXSMA - CUP E63C22002180006.
A.~T. and G.~S. acknowledge financial support through the project with grant doi: 10.55776/I5487 by the Austrian Science Fund (FWF) and project Nr.~468199700 by the DFG, respectively as well as project with grant doi: 10.55776/I5868 (P1) as well as 10.55776/KIN256372 and project P5 of the FOR 5249 [QUAST] by the DFG Nr.~449872909. 
D.C. gratefully acknowledges the support provided by the Leverhulme Trust, as well as additional support from the UKRI Horizon Europe Guarantee Grant No. EP/Z002419/1 (for an ERC Consolidator Grant to Siddharth A. Parameswaran).
We also thank the Gauss Center for Supercomputing e.V. (www.gauss-center.eu) for funding this project by providing computing time on the GCS Supercomputer SuperMUC at Leibniz Supercomputing Center (www.lrz.de). 
\end{acknowledgments}

\section*{Data availability}
The data that support the findings of this article are publicly available \cite{data}

\begin{appendix}

\newpage
\section{Scattering phase shift}
\label{App:Scattering_phase_shift}
Following the calculations in \cite{Coleman_2015} for the scattering phase shift between the localized impurity fermion with the electron bath, we look at our Anderson impurity Hamiltonian equation \eqref{eq:general_AIM_Hamiltonian} for a single impurity site without interactions ($U\,=\,0$). The tunneling between the impurity and the bath is described by the hybridization function:
\begin{equation}
    \Delta(\omega)\,=\,\int^{\mathcal{D}}_\mathcal{-D}\,\text{d}\varepsilon\,\left[-\frac{1}{\pi}\,\frac{\text{Im}\left[\Delta(\varepsilon)\right]}{\omega + i 0^{+}- \varepsilon}\right]\,.
\end{equation}
As in the rest of this work, with $\mathcal{D}$ and $\mu$ we denote the frequency cutoff and the chemical potential, respectively while $0^{+}$ is a small imaginary part. The retarded noninteracting impurity Green's function can then be written as
\begin{equation}
    G_0(\omega)\,=\,\frac{1}{\omega\,+\,i\,0^{+} \,-\,\epsilon_{f}\,-\,\Delta(\omega)}\,,
    \label{eq-suppl:Green}
\end{equation}
where $\epsilon_{f}$ is set to zero for the system under consideration. The corresponding density of states on the impurity reads
\begin{equation}
    \text{DOS}(\omega)\,=\,-\,\frac{1}{\pi}\,\text{Im}\left[G_0(\omega)\right]\,.
    \label{eq-suppl:DOS}
\end{equation}
As the next step, we use two different forms of the scattering $T$ matrix:\\
(i) Defining the $T$-matrix as bath electrons scattering at the impurity depending on the hybridization strength $V$ and moving back into the bath, it yields
\begin{equation}
    T(\omega)\,=\,V^2\,G_0(\omega)\,.
\end{equation}
(ii) The second relation for the $T$ matrix comes from the definition of the $S$ matrix by the scattering phase shift $\xi(\omega)$ as
\begin{equation}
    S(\omega)\,=\,e^{2\,i\,\xi(\omega)}\,.
\end{equation}
This $S$ matrix is connected to the $T$ matrix by the following formula:
\begin{equation}
    S(\omega)\,=\,1\,-\,2\,\pi\,i\,\rho(\omega)\,T(\omega)\,,
\end{equation}
with $\rho(\omega)\,=\,-\frac{1}{\pi\,V^2}\,\text{Im}\left[\Delta(\omega)\right]$ being the density of states of the bath fermions. Solving these equations for the $T$ matrix leads to
\begin{equation}
    T(\omega)\,=\,-\frac{1}{\pi\,\rho(\omega)}\,\frac{1}{\cot{\left(\xi(\omega)\right)}\,-\,i}
\end{equation}

Comparing both expressions for the $T$ matrix returns the scattering phase shift depending on the hybridization function:
\begin{widetext}
\begin{equation}\label{eq:scatt_phase_shift}
    \xi(\omega)\,=\,\text{arccot}\left(\frac{\omega\,-\,\text{Re}\left[\Delta(\omega)\right]}{\text{Im}\left[\Delta(\omega)\right]}\right)\,=\,\begin{cases} \text{arctan}\left(\frac{\text{Im}\left[\Delta(\omega)\right]}{\omega\,-\,\text{Re}\left[\Delta(\omega)\right]}\right),\quad \text{if} \quad \text{arg}(\text{arctan})\,\ge\,0 \\ \text{arctan}\left(\frac{\text{Im}\left[\Delta(\omega)\right]}{\omega\,-\,\text{Re}\left[\Delta(\omega)\right]}\right)\,+\,\pi,\quad \text{if} \quad \text{arg}(\text{arctan})\,<\,0\end{cases}\,.
\end{equation}
\end{widetext}
At small $\omega$ the real part $\Delta(\omega)$ goes as $\simeq \frac{\alpha V^2}{\omega}$ for any small $\alpha \neq 0$ while the imaginary part is diverging and always negative. Therefore, Eq.~(\ref{eq:scatt_phase_shift}) gives a jump from $\xi = \pi$ at 
$\omega = 0^-$ to $\xi = 0$ at $\omega = 0^+$. As the singular part of  $\Delta(\omega)$ dominates, then the impurity spectral function develops two peaks  at $\omega=\omega^*\simeq \sqrt{\alpha} V$. 
On the other hand, at energies $\omega\simeq\omega^*$ only the regular part of $\Delta$ contributes and the phase shift is restricted to values around $\pi/2$. 
This suggests that we consider $\omega^*$ as a specific energy cutoff set by $\alpha$ above which the system recovers the Kondo physics. 
Our RG analysis, together with the susceptibility of the main
text, supports the transition from non-Kondo to local-moment physics for thermodynamic quantities, in the limit of small $\alpha$ at finite temperature. 

\begin{figure}[]
    \centering
    \includegraphics[width=0.49\textwidth]{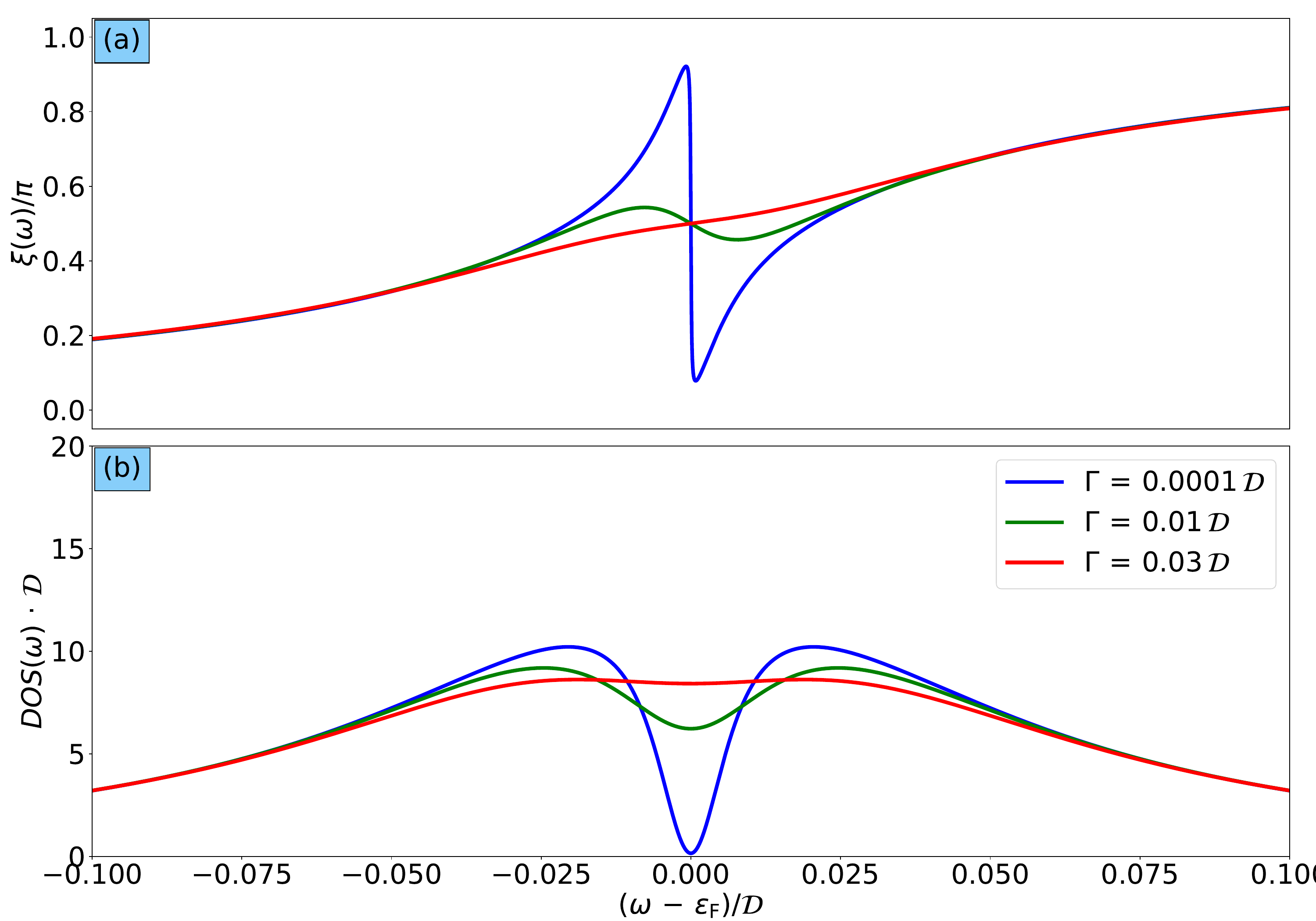}
    \caption{\justifying Scattering phase shift (a) and impurity density of states (b) for the noninteracting ($U=0$) single-site Anderson impurity model at $V=0.2\,\mathcal{D}$ with an additional small broadening $\Gamma$ of the $\delta$ peak (of weight $\alpha=0.01$) in the hybridization function.}
\label{fig:scattering_phase_shift_and_impurity_DOS_Lorentzian}
\end{figure}

In particular, considering the RG flow at the energy scale set by the temperature, would amount to adding an effective broadening $\Gamma$ to the $\delta$ peak of the hybridization function. Hence, such finite-temperature situation can be modeled as shown in Fig.~\ref{fig:scattering_phase_shift_and_impurity_DOS_Lorentzian}:
For small $\alpha$, the system is able to overcome the phase-shift jump at the
Fermi level due to the broadening of the $\delta$-peak when $\Gamma \simeq \omega^*$.
In this context, we note that additional sources of the intrinsic broadening $\Gamma$ would also be self-energy effects entering from the DMFT self-consistency at finite temperature. Their quantification is obviously system-dependent but Fig.~\ref{fig:scattering_phase_shift_and_impurity_DOS_Lorentzian} shows that even a tiny value of $\Gamma$ yields a substantial smoothening of the $\pi$-jump.

\section{Limit of high temperature and large interaction for the effective moment}
In the infinite temperature limit $T\,\rightarrow{\infty}$ all eigenstates $|\,i\big>$ of a system are equally distributed following the Boltzmann distribution: ${e^{-\beta\,\epsilon_i}\,=\,1\,\,\,\,\forall\,\epsilon_i}$, with the eigenenergies $\epsilon_i$ shifted by the half filling chemical potential. We begin with the definition of the expectation value for the squared $z$-spin operator $\big<\hat{S}^2_z\big>$:
\begin{equation}
    \big<\hat{S}^2_z\big>\,=\,\frac{1}{Z}\,\sum_{i}\,e^{-\beta\,\epsilon_i}\,\big<i\,|\hat{S}^2_z|\,i\big>.
\end{equation}
The corresponding partition function $Z$ reads
\begin{equation}
    Z\,=\,\sum_{i}\,1\,=\,2^{2\,N_{\text{orb}}}\,,
\end{equation}
with the number of orbitals on the impurity site $N_{\text{orb}}$. Using the representation $\hat{S}_z\,=\,\frac{\hat{n}_{\uparrow}\,-\,\hat{n}_{\downarrow}}{2}$ for the $z$-spin operator, with $n_\uparrow$ the number of spin-up and $n_\downarrow$ the number of spin-down electrons and taking into account all possibilities for up and down electrons on the impurity site, we get the following representation,
\begin{equation}
\begin{split}
        \sum_{i}\big<i|\hat{S}^2_z|i\big>&=\frac{1}{4}\sum^{N_{\text{orb}}}_{n_\uparrow,n_\downarrow=0}\binom{N_{\text{orb}}}{n_\uparrow}\binom{N_{\text{orb}}}{n_\downarrow}\left(n_\uparrow-n_\downarrow\right)^2=\\&=2^{(2\,N_{\text{orb}}-1)}\,\frac{N_{\text{orb}}}{4}\,=\,2^{(2\,N_{\text{orb}}-3)}\,N_{\text{orb}}
\end{split}
\end{equation}
and for the expectation value of the squared spin-$z$ operator,
\begin{equation}
    \big<\hat{S}^2_z\big>(T\,\rightarrow{\infty})\,=\,\frac{1}{Z}\,\left(2^{(2\,N_{\text{orb}}\,-\,3)}\,N_{\text{orb}}\right)\,=\,\frac{N_{\text{orb}}}{8}\,.
\end{equation}
Thus the overall local spin susceptibility has the form
\begin{equation}
    \chi^{\text{spin}}_{\text{zz}}(T\,\rightarrow{\infty})\,=\,\frac{g^2\,\big<\hat{S}^2_z\big>}{T}\,=\,\frac{N_{\text{orb}}}{2\,T}\,.
\end{equation}
Comparing this with a Curie behavior including an effective moment $\mu_\text{eff}$,
\begin{equation}
    \chi^{\text{spin}}_{\text{zz}}\,=\,\frac{\mu^2_\text{eff}}{3\,T}\,,
\end{equation}
yields
\begin{equation}
    \mu_\text{eff}(T\,\rightarrow{\infty})\,=\,\sqrt{\frac{3\,N_\text{orb}}{2}}\,.
\end{equation}

\section{High-temperature expansion for the effective moment}
To investigate the large temperature expansion of the effective moment $\mu_\text{eff}$ we assume the Hamiltonian for the Kondo dimer in Eq.(\ref{eq:general_AIM_Hamiltonian}) for $\alpha=1$ perturbed by an external magnetic field $h$:
\begin{equation}
    \hat{H}_\text{perturbed}\,=\,\hat{H}\,-\,h\,\left(\hat{f}^\dagger_{\uparrow}\,\hat{f}_{\uparrow}\,-\,\hat{f}^\dagger_{\downarrow}\,\hat{f}_{\downarrow}\right)\,.
\end{equation}
Since the one- and three-particle subspaces have the highest contributions to the local spin susceptibility at large temperatures, we focus only on these cases. The Hamiltonian of the one- (three-) particle subspace reads
\begin{equation}
\begin{split}
    \hat{H}_{1 (3)}&=\begin{pmatrix}0 && (-)V && 0 && 0\\ (-)V && \frac{U}{2}-h && 0 && 0\\ 0 && 0 && 0 && (-)V\\ 0 && 0 && (-)V && \frac{U}{2}+h\end{pmatrix}\\
    &=\begin{pmatrix}n_0\mathbb{1}\pm n_1\sigma_x+n_3\sigma_z && 0\\ 0 && n'_0\mathbb{1}\pm n'_1\sigma_x+n'_3\sigma_z\end{pmatrix}\,,
\end{split}
\end{equation}
with the eigenvalues
\begin{equation}
\begin{split}
        \epsilon_{1,\,2}&=n_0\,\pm\,\vert{\mathbf{n}}\vert{}\\
        \epsilon_{3,\,4}&=n'_0\,\pm\,\vert{\mathbf{n}'}\vert{}\,.
\end{split}
\end{equation}
\begin{figure}[t!]
    \centering \includegraphics[width=0.49\textwidth]{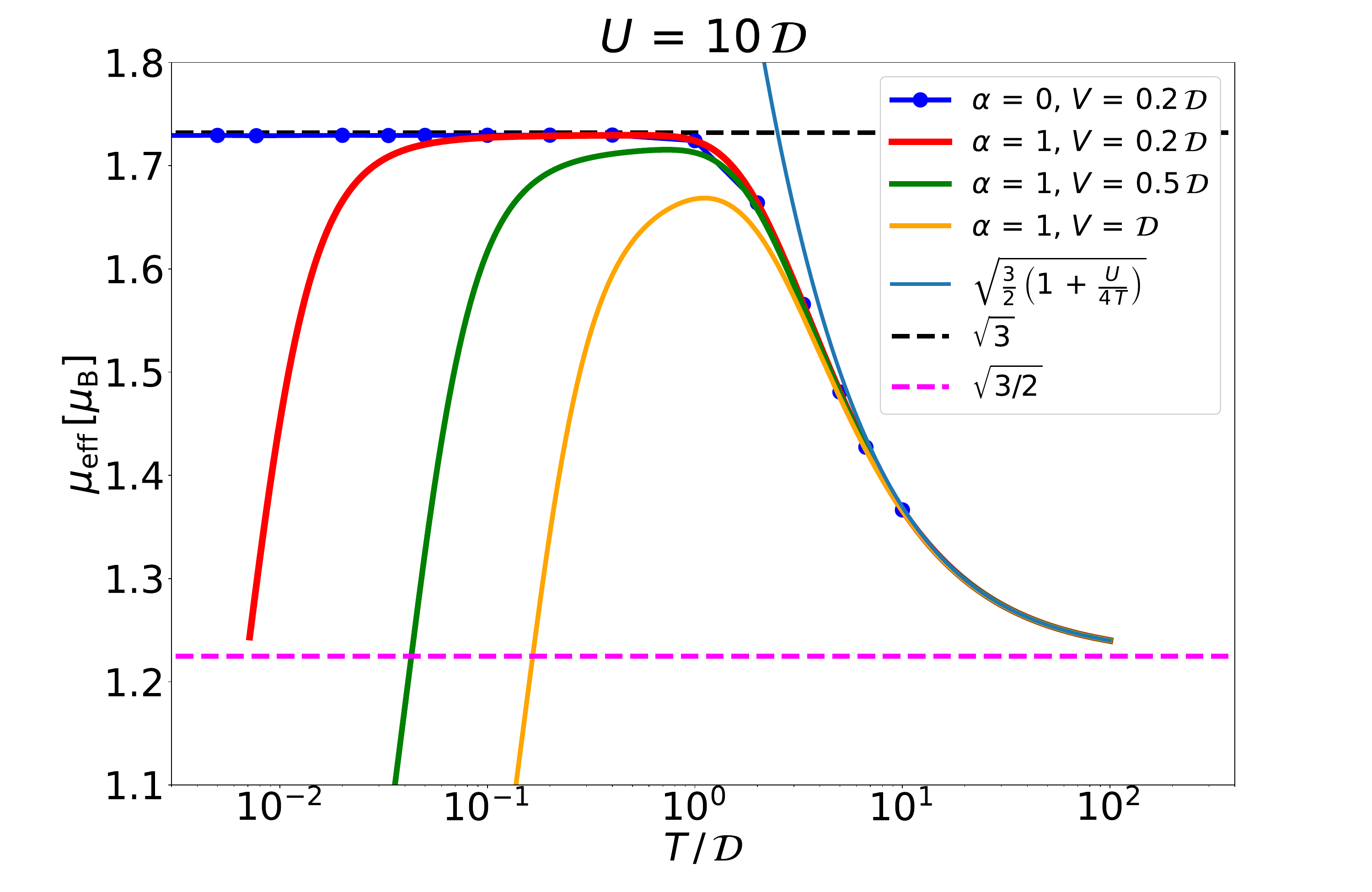}
    \caption{\justifying Comparison of the curves for analytic calculations of a Kondo dimer with interaction $U\,=\,10\,\mathcal{D}$ at several hybridizations [$V\,=\,0.2\,\mathcal{D}$ (red), $V\,=\,0.5\,\mathcal{D}$ (green), $V\,=\,\mathcal{D}$ (orange)]. The plateau of the large $U$ limit for the effective moment is not reached at larger hybridizations. Purple: Large temperature expansion of analytic calculations within perturbation theory. Blue: $\mu_\text{eff}$ for a boxlike hybridization function at $U\,=\,10\,\mathcal{D}$ and $V\,=\,0.2\,\mathcal{D}$. Here, the curve reaches the maximum effective moment over a long temperature range.}
    \label{fig:Comparison_mu_eff_dif_V}
\end{figure}
Here, we renamed the prefactors of the Pauli matrices as $\mathbf{n}^{(}{'^{)}}\,=\,\left(n_1{^{(}{'^{)}}},\,\,n_2{^{(}{'^{)}}},\,\,n_3{^{(}{'^{)}}}\right)^\text{T}$ and of the unity matrix as $n_0{^{(}{'^{)}}}$.
The partition functions of the two $2\times2$ matrix blocks are
\begin{equation}
\begin{split}
    Z_{1 (3)}\,&=\,e^{-\beta\,\epsilon_1}\,+\,e^{-\beta\,\epsilon_2}\\
    Z'_{1 (3)}\,&=\,e^{-\beta\,\epsilon_3}\,+\,e^{-\beta\,\epsilon_4}\,.
\end{split}
\end{equation}
As the next step we determine the magnetization in the canonical ensemble
\begin{equation}
    M\,=\,\big<\,\hat{S}_z\,\big>\,=\,\frac{\text{Tr}\left[e^{-\beta\,\hat{H}_\text{perturbed}}\,\hat{S}_z\right]}{\text{Tr}\left[e^{-\beta\,\hat{H}_\text{perturbed}}\right]}\,,
\end{equation}
by introducing the $z$-spin operator $\hat{S}_z$:
\begin{equation}
    \hat{S}_z=\begin{pmatrix} 1&0\\0&-1\end{pmatrix}\otimes\begin{pmatrix} 0&0\\0&1\end{pmatrix}\,=\,\begin{pmatrix}\frac{1}{2}\mathbb{1}-\frac{1}{2}\sigma_z && 0\\ 0 && -\frac{1}{2}\mathbb{1}+\frac{1}{2}\sigma_z\end{pmatrix}\,.
\end{equation}
Using the properties of the Pauli matrices, the exponential in the magnetization can be written as [${c^{(}{'^{)}}=\cosh{\left(\beta\,\vert{\mathbf{n}^{(}{'^{)}}}\vert{}\right)}, s^{(}{'^{)}}=\sinh{\left(\beta\,\vert{\mathbf{n}^{(}{'^{)}}}\vert{}\right)}}$]:
\begin{equation}
    e^{-\beta\,\hat{H}^{2\times2}_\text{perturbed}}\,=\,e^{-\beta\,n_0{^{(}{'^{)}}}}\,\left(c^{(}{'^{)}}\,\mathbb{1}\,-\,s^{(}{'^{)}}\,\frac{\mathbf{n}^{(}{'^{)}}}{\vert{\mathbf{n}^{(}{'^{)}}}\vert{}}\,\cdot\,\boldsymbol{\sigma}\right)\,,
\end{equation}
for the respective $2\times2$ blocks and both subspaces.
With these expressions we are able to establish the overall magnetization in the one- and three-particle sectors:
\begin{equation}
\begin{split}
        M_{1 (3)}&=\frac{1}{Z_{1 (3)}+Z'_{1 (3)}}\\
        &\times\left[e^{-\beta n_0}\left(c+s\cdot\frac{n_3}{\vert{\mathbf{n}}\vert{}}\right)-e^{-\beta n'_0}\left(c'+s'\cdot\frac{n'_3}{\vert{\mathbf{n}'}\vert{}}\right)\right].
\end{split}
\end{equation}
Since we are interested in the limit of high temperatures, we expand the magnetization around $\beta\,=\,0$ up to second order and by performing the derivative for the magnetic field $h$, we arrive at the high temperature expansion of the local spin susceptibility,
\begin{equation}
    \chi^{\text{spin}}_{zz}(T\,\rightarrow{\infty})\,=\,\frac{\partial M_{1 (3)}(T\,\rightarrow{\infty})}{\partial h}\,\approx\,\beta\,\left(\frac{1}{2}\,+\,\beta\,\frac{U}{8}\right)
\end{equation}
and with the definition $\mu_\text{eff}\,=\,\sqrt{3\,T\,\chi^{\text{spin}}_{zz}}$ we find the effective moment at high temperatures up to first order in $\beta$:
\begin{equation}
    \mu_\text{eff}(T\rightarrow{\infty})=\sqrt{\frac{3}{2}\left(1+\frac{U}{4T}\right)}\approx\sqrt{\frac{3}{2}}\left(1+\frac{U}{8T}\right).
\end{equation}
Figure\,\ref{fig:Comparison_mu_eff_dif_V} shows the high-temperature expansion for the effective moment (purple) together with curves of analytic results for a Kondo dimer at an interaction of $U\,=\,10\,\mathcal{D}$ for different values of the hybridization $V$. Increasing $V$ in relation to $U$, the effective moment $\mu_\text{eff}$ no longer reaches its maximum at the plateau $\mu_\text{eff}(U\,\rightarrow{\infty})\,=\,\sqrt{3}$, since the formation of a singlet ground state is favored. 

\section{Hybridization function for Dirac semimetals and for the THF model}
\label{app:Hybridization_THF}
In the following we derive the hybridization function for two different lattice models obtained through the projection onto a single-site Anderson impurity model: we investigate two single-orbital Hamiltonians with linear energy versus momentum dispersion in three and two spatial dimensions, respectively. For both models we assume half-filling and particle-hole symmetry. Further, as in the rest of the paper, we do not explicitly include the back effect of a local self-energy, which is expected to be relevant only in those cases in which the DMFT self-consistency qualitatively changes the low-energy structure, such as for instance in Mott phases.

\begin{figure}
    \centering
    \includegraphics[width=0.49\textwidth]{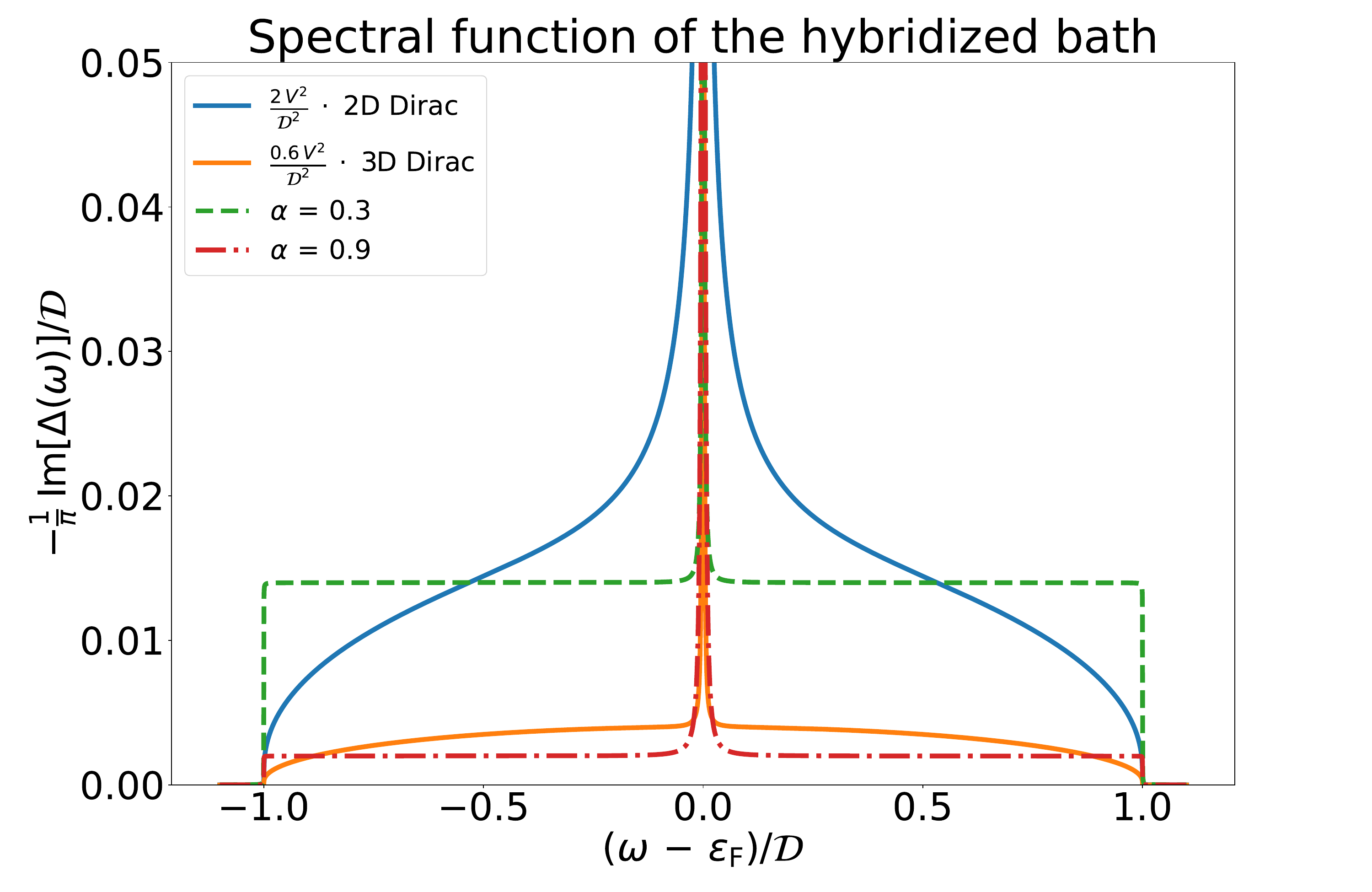}
    \caption{\justifying Imaginary part of hybridization functions $-\frac{1}{\pi}\,\text{Im}[\Delta(\omega)]/\mathcal{D}$ for 2D and 3D Dirac systems (blue and orange) together with the model of the main text combining a boxlike hybridization with a central $\delta$-peak for two possible values of $\alpha$ ($\alpha\,=\,0.3$ in green and $\alpha\,=\,0.9$ in red) with a small imaginary part $0^{+}$ added to the frequency. The hybridization function for the 2D Dirac cone is multiplied by the factor $\frac{2\,V^2}{\mathcal{D}^2}$ and for the 3D Dirac semimetal it is multiplied by $\frac{0.6\,V^2}{\mathcal{D}^2}$, such that they are normalized to $V^2$. Parameters: $V\,=\,0.2\,\mathcal{D}$, $0^{+}\,=\,0.0001\,\mathcal{D}$.}
    \label{fig:Comparison_2d_3d_flat_plus_delta}
\end{figure}

For the three-dimensional (3D) and two-dimensional (2D) Dirac semimetals we assume an energy cutoff $\mathcal{D}$. In both cases the dispersion reads ${\epsilon_{\vert{\mathbf{k}}\vert{}}\,=\,\vert{\mathbf{k}}\vert{}}$. Starting with the three-dimensional Dirac cone, the corresponding $-\frac{1}{\pi}\,\text{Im}\left[G_0(\epsilon)\right]$ becomes
\begin{equation}
    -\frac{1}{\pi}\,\text{Im}\left[G_0(\epsilon)\right]\,\propto\,\int^{\infty}_{-\infty}\,\text{d}^3 \mathbf{k}\,\delta(\epsilon\,-\,\epsilon_{\vert{\mathbf{k}}\vert{}})\,\propto\,\epsilon^2
\end{equation}
With the condition of a normalized $-\frac{1}{\pi}\,\text{Im}\left[G_0(\epsilon)\right]$, this yields
\begin{equation}
    -\frac{1}{\pi}\,\text{Im}\left[G_0(\epsilon)\right]\,=\,\frac{3}{2\,\mathcal{D}^3}\,\epsilon^2
\end{equation}
Using this result, we are able to calculate the noninteracting Green's function $G_0$,
\begin{equation}
\begin{split}
        G_0(\omega)&=\int^{\infty}_{-\infty}\text{d}\epsilon\frac{-\frac{1}{\pi}\text{Im}\left[G_0(\epsilon)\right]}{\omega+i 0^{+}-\epsilon}=-\frac{3\left(\omega+i 0^{+}\right)}{2\mathcal{D}^3}\times\\&\times\left[2\mathcal{D}+\left(\omega+i 0^{+}\right)\ln{\left(\frac{\omega+i 0^{+}-\mathcal{D}}{\omega+i 0^{+}+\mathcal{D}}\right)}\right],
\end{split}
\end{equation}
where $0^{+}$ is a small imaginary part. If we compute the real and the imaginary part for the logarithm of the previous expression and take the limit where $\omega$ and $0^{+}$ are small, we get $\ln{\left(\frac{\omega\,+\,i\,0^{+}\,-\,\mathcal{D}}{\omega\,+\,i\,0^{+}\,+\,\mathcal{D}}\right)}\,\approx\,-i\,\frac{2\,0^{+}}{\mathcal{D}}\,+\,i\,\pi$ and find
\begin{equation}
\begin{split}
        G_0(\omega)&=G'_0+iG''_0\\&\approx-\frac{3}{\mathcal{D}^2}\left(1+\frac{\pi 0^{+}}{\mathcal{D}}\right)\omega-\\&-i\frac{3}{\mathcal{D}^2}\left(0^{+}+\frac{\pi}{2\mathcal{D}}\omega^2-\frac{\pi}{2\mathcal{D}}\left(0^{+}\right)^2\right).
\end{split}
\end{equation}
To determine the hybridization function we use the formula
\begin{equation}
\begin{split}
        \Delta(\omega)\,&=\,\omega\,+\,i\,0^{+}\,-\,\left[G_0(\omega)\right]^{-1}\\&=\,\frac{\omega\,\left(G'^2_0+G''^2_0\right)+i 0^{+}\left(G'^2_0+G''^2_0\right)-G'_0+iG''_0}{G'^2_0+G''^2_0}\,,
\end{split}
\end{equation}
with $\mathcal{H}_\mathrm{loc}\,=\,0$. Inserting the previous approximations together with the limits $0^{+}\,\rightarrow{0}$ and $\omega\,\rightarrow{0}$ yields
\begin{equation}
\begin{split}
        -\frac{1}{\pi}\,\text{Im}\left[\Delta(\omega)\right]\,&=\,-\frac{1}{\pi}\,\left[0^{+}\,+\,\frac{G''_0}{G'^2_0\,+\,G''^2_0}\right]\\ &\approx\,\frac{\mathcal{D}^2}{3}\,\delta(\omega)\,+\,\frac{\mathcal{D}}{6\,\pi}\,,
\end{split}
\end{equation}
describing a Dirac $\delta$ function together with a constant background.
\\\\
In order to find the correct behavior of the peaked structure for the hybridization of the $f$ bands in TBG at low frequencies, we proceed by analyzing the case of a 2D Dirac system rather than the previously calculated 3D behavior. Thus the density of states has the form
\begin{equation}
    -\frac{1}{\pi}\,\text{Im}\left[G_0(\epsilon)\right]\,\propto\,\int^{\infty}_{-\infty}\,\text{d}^2 \mathbf{k}\,\delta(\epsilon\,-\,\epsilon_{\vert{\mathbf{k}}\vert{}})\,\propto\,\vert{\epsilon}\vert{}\,,
\end{equation}
with its normalized expression:
\begin{equation}
    -\frac{1}{\pi}\,\text{Im}\left[G_0(\epsilon)\right]\,=\,\frac{1}{\mathcal{D}^2}\,\vert{\epsilon}\vert{}\,.
\end{equation}
Following the procedure in three dimensions, the noninteracting Green's function $G_0$ reads
\begin{equation}
    G_0(\omega)\,=\,\frac{\left(\omega\,+\,i\,0^{+}\right)}{\mathcal{D}^2}\,\ln{\left(\frac{\left(\omega\,+\,i\,0^{+}\right)^2}{\left(\omega\,+\,i\,0^{+}\right)^2\,-\,\mathcal{D}^2}\right)}\,.
\end{equation}
By taking the limits $\omega\,\rightarrow{0}$ and $0^{+}\,\rightarrow{0}$ the logarithm becomes $\ln{\left(\frac{\left(\omega\,+\,i\,0^{+}\right)^2}{\left(\omega\,+\,i\,0^{+}\right)^2\,-\,\mathcal{D}^2}\right)}\,\approx\,x\,+\,i\,y\,-\,i\,z$, with $x\,=\,\ln{\left(\frac{\omega^2\,+\,\left(0^{+}\right)^2}{\mathcal{D}^2}\right)}$, $y\,=\,\arctan{\left(-\frac{2\,\omega\,0^{+}}{\left(0^{+}\right)^2\,-\,\omega^2}\right)}$ and $z\,=\,\pi\,\text{sig}(\omega)$, yielding\\
\begin{equation}
    G_0(\omega)\,=\,G'_0\,+\,i\,G''_0\,\approx\,\frac{\left(\omega + i 0^{+} \right)}{\mathcal{D}^2}\,\left(x\,+\,i\,y\,-\,i\,z\right).
\end{equation}
\noindent This leads to the imaginary part of the hybridization functions for $\omega\,\rightarrow{0}$ and $0^{+}\,\rightarrow{0}$:
\begin{equation}
\begin{split}
        -\frac{1}{\pi}\,\text{Im}\left[\Delta(\omega)\right]\,&=\,-\frac{1}{\pi}\,\left[0^{+}\,+\,\frac{G''_0}{G'^2_0\,+\,G''^2_0}\right]\\&\approx\,\mathcal{D}^2\,\text{sig}(\omega)\,\frac{1}{\omega\,\left(4\,\ln{\left(\frac{\omega}{\mathcal{D}}\right)}^2\,+\,\pi^2\right)}\,.
\end{split}
\end{equation}
Figure\,\ref{fig:Comparison_2d_3d_flat_plus_delta} compares the quantity $-\frac{1}{\pi}\,\text{Im}[\Delta(\omega)]/\mathcal{D}$ for a two- and a three-dimensional Dirac semimetal and the toy model of the main text for $\alpha\,=\,0.3$ and $\alpha\,=\,0.9$. Here, in all cases the full noninteracting Green's function was taken to calculate the hybridization functions. For the 2D Dirac cone the curve was multiplied by $\frac{2\,V^2}{\mathcal{D}^2}$ and for the 3D Dirac semimetal it was multiplied by $\frac{0.6\,V^2}{\mathcal{D}^2}$ to be normalized to $V^2$.

\section{Derivation of the RG flow equations}
\begin{figure*}[t!]
    \centering \includegraphics[width=0.99\textwidth]{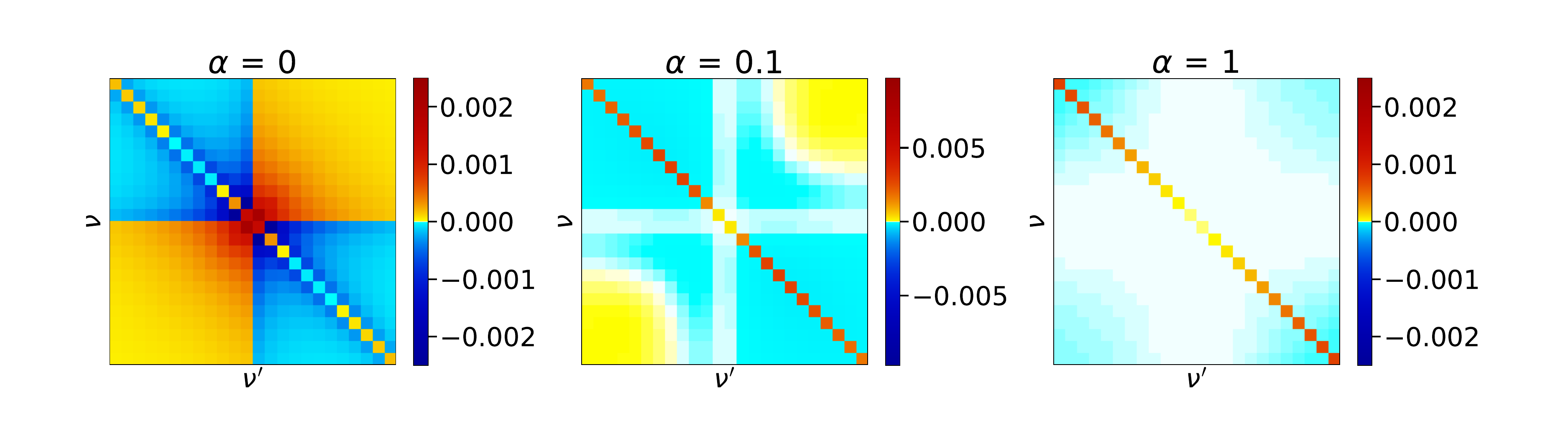}
    \caption{\justifying Matrices of the generalized static charge susceptibility at $T\,\approx\,0.0017\,\mathcal{D}$ for $\alpha\,=\,0$ (left), $\alpha\,=\,0.1$ (center) and $\alpha\,=\,1$ (right). The onionlike structure of the continuous hybridization ($\alpha\,=\,0$) vanishes due to the weight of the $\delta$-peak in the hybridization function leading to a minimum at small Matsubara frequencies in the center of the matrix.}
    \label{fig:colorplots_general_chi}
\end{figure*}
Our RG calculations are based on the Kondo Hamiltonian Eq.~(\ref{eq:Kondo_Hamiltonian}). Here, we introduced the Abrikosov pseudospin ${\mathbf{S}\,=\,\hat{f}^\dagger_{\sigma}\,\frac{\boldsymbol{\sigma}_{\sigma,\sigma'}}{2}\,\hat{f}_{\sigma'}}$ with the constraint $\sum_\sigma\,\hat{f}^\dagger_{\sigma}\,\hat{f}_\sigma\,=\,1$. The corresponding action can be written as
\begin{widetext}
\begin{equation}
\begin{split}
    \mathcal{S}\,=\,\int^{\beta}_{0}\,\text{d}\tau\,\Bigg[&\bar{f}_\sigma\,\partial_\tau\,f_\sigma\,+\,\int_{-\Lambda_0}^{\Lambda_0}\,\text{d}k\,\bar{c}_\sigma(k,\,\tau)\,\left(\partial_\tau\,-\,k\right)\,c_\sigma(k,\,\tau)\,-\,J\,\left(\bar{f}_\sigma\,\frac{\boldsymbol{\sigma}_{\sigma,\sigma'}}{2}\,f_{\sigma'}\right)\,\cdot\,\left(\bar{c}_\sigma(0)\,\frac{\boldsymbol{\tau}_{\sigma,\sigma'}}{2}\,c_{\sigma'}(0)\right)\\
    &+\,\bar{d}_{\sigma}\,\partial_\tau\,d_{\sigma}\,-\,K\,\left(\bar{f}_{\sigma}\,\frac{\boldsymbol{\sigma}_{\sigma,\sigma'}}{2}\,f_{\sigma'}\right)\,\cdot\,\left(\bar{d}_{\sigma}\,\frac{\boldsymbol{\tau}_{\sigma,\sigma'}}{2}\,d_{\sigma'}\right)\\
    &-\,M\,\left(\bar{f}_{\sigma}\,\frac{\boldsymbol{\sigma}_{\sigma,\sigma'}}{2}\,f_{\sigma'}\right)\,\cdot\,\left(\bar{c}_\sigma(0)\,\frac{\boldsymbol{\tau}_{\sigma,\sigma'}}{2}\,d_{\sigma'}\,+\,\bar{d}_\sigma\,\frac{\boldsymbol{\tau}_{\sigma,\sigma'}}{2}\,c_{\sigma'}(0)\right)\Bigg]\,,
\end{split}
\end{equation}
\end{widetext}
subject to the constraint on the $f$-fermions. Note that we introduced an energy cutoff $\Lambda$, and the fermions are represented by Grassmann variables. The Kondo couplings are related to that of the Anderson model in Eq.~\eqref{eq:general_AIM_Hamiltonian} of the main text as $J=(1-\alpha) J_0$, $K=\alpha J_0$ and $M=\sqrt{\alpha(1-\alpha)} J_0$ with $J_0 = 2|V|^2 (1/|\epsilon_{f}| + 1/|\epsilon_{f}+U|)$, and we recall that $\alpha$ denotes the weight fraction of the peaked contribution of the Anderson-model hybridization function.
The scaling dimensions of the three coupling constants can be read off as
\begin{align}
    \text{dim}[J]\,=\,0\,,\,\,\,\,\,\,\,\,\,\,\text{dim}[K]\,=\,-1\,,\,\,\,\,\,\,\,\,\,\,
    \text{dim}[M]\,=\,-\frac{1}{2}\,.
\end{align}
The perturbative RG flow equations for the dimensionless running couplings $j$, $k$ and $m$ are given in Eq.~\eqref{eq:flow_equations}. Their linear terms are determined by the scaling dimension of the couplings; further terms describe contributions obtained by integrating out high-energy bath degrees of freedom. In our problem there are no $d$ fermions to be integrated out. The Kondo-only beta function arising for the $j$ coupling is known from previous RG calculations \cite{PhysRevB.70.214427}; perturbative renormalizations for $k$ and $m$ arise due to the presence of the mixed $m$ coupling.

\section{Generalized charge susceptibility}
In the main text we analyzed the contribution of the bubble and vertex corrections to the local spin susceptibility depending on the weight of the central $\delta$ peak controlled by the parameter $\alpha$. Here, we are interested in the behavior of the two-particle generalized charge susceptibility $\chi^{\text{charge}}_{\nu\nu'\omega}$ with respect to the chosen value of $\alpha$. This quantity including the Matsubara frequencies $\nu$ and $\nu'$ considers also two-particle interactions and contributes to the impurity charge susceptibility of the main text after summation over the frequencies. The general static form ($\omega\,=\,0$) of this quantity reads

\begin{equation}
    \chi^{\text{charge}}_{\nu\nu'}\,=\,\chi^{\uparrow\uparrow}_{\nu\nu'}\,+\,\chi^{\uparrow\downarrow}_{\nu\nu'}
\end{equation}
with
\begin{widetext}
    \begin{equation}
        \chi^{\sigma\sigma'}_{\nu\nu'}=\int^{\beta}_0\text{d}\tau_1\text{d}\tau_2\text{d}\tau_3e^{-i\nu\tau_1}e^{i\nu\tau_2}e^{-i\nu'\tau_3}\big[\big<T_\tau\hat{f}^\dagger_{\sigma}(\tau_1)\hat{f}_{\sigma}(\tau_2)\hat{f}^\dagger_{\sigma'}(\tau_3)\hat{f}_{\sigma'}(0)\big>-\big<T_\tau\hat{f}^\dagger_{\sigma}(\tau_1)\hat{f}_{\sigma}(\tau_2)\big>\big<T_\tau\hat{f}^\dagger_{\sigma'}(\tau_3)\hat{f}_{\sigma'}(0)\big>\big].
\end{equation}
\end{widetext}
The matrix of the generalized charge susceptibility $\chi^{\text{charge}}_{\nu\nu'}$ is plotted in Fig.\,\ref{fig:colorplots_general_chi} for different values of $\alpha$. For $\alpha\,=\,0$ we are able to reproduce an onionlike structure, discussed in \cite{PhysRevLett.126.056403}, with a perturbative region at larger frequencies, a local moment regime at intermediate frequencies and a region at small frequencies, where the local moment is screened. The screening leads to a relaxation of the generalized charge susceptibility. Increasing $\alpha$ destroys the local moment together with the screening features, yielding two regimes: (i) a perturbative region at large frequencies and (ii) a region at smaller frequencies, where the diagonal is suppressed. Off-diagonal entries also decrease with $\alpha$. The suppression of $\chi^{\text{charge}}_{\nu,\,\nu'}$ depends on the weight of the hybridization function around the Fermi level.

\end{appendix}

\bibliography{bibliography}

\end{document}